\documentclass[3p,final,10pt]{elsarticle}
\usepackage{ifxetex,ifluatex}
\if\ifxetex T\else\ifluatex T\else F\fi\fi T%
  \usepackage{fontspec}
\else
  \usepackage[T1]{fontenc}
  \usepackage[utf8]{inputenc}
  \usepackage{lmodern}
\fi

\usepackage{amsmath}
\usepackage{amssymb}
\usepackage{graphicx}
\usepackage{subfig}
\usepackage{booktabs} 
\usepackage{newtxtext}%
\usepackage{newtxmath}%
\usepackage{arydshln}
\usepackage[final,nopatch=footnote]{microtype}
\usepackage{todonotes}
\usepackage{lineno}
\usepackage[hang,flushmargin]{footmisc}
\usepackage{tcolorbox} 
\usepackage{xcolor} 
\usepackage{hyperref}
\hypersetup{colorlinks=true, linkcolor=blue, citecolor=blue, urlcolor=blue}
\usepackage{xurl}
\newcommand{\Alphaset}{\{\boldsymbol{\alpha}\}}
\newcommand{\Alphasett}{\{\boldsymbol{\alpha}\}}
\newcommand{\Alphadotset}{\{\dot{\boldsymbol{\alpha}}\}}

\newcommand{\Chiset}{\{\boldsymbol{\chi}\}}
\newcommand{\Chidotset}{\{\dot{\boldsymbol{\chi}}\}}
\newcommand{\Alphafab}{\boldsymbol{\alpha}^{\text{fab}}}
\newcommand{\Alphastruct}{\boldsymbol{\alpha}^{\text{str}}}
\newcommand{\AlphaStruct}{\boldsymbol{\alpha}^{\text{Str}}}
\newcommand{\Alphadotfab}{\dot{\boldsymbol{\alpha}}^{\text{fab}}}
\newcommand{\Alphadotstruct}{\dot{\boldsymbol{\alpha}}^{\text{str}}}
\newcommand{\AlphadotStruct}{\dot{\boldsymbol{\alpha}}^{\text{Str}}}
\newcommand{\Alphadotpre}{\dot{\boldsymbol{\alpha}}^{\text{pre}}}
\newcommand{\Alphapre}{\boldsymbol{\alpha}^{\text{pre}}}

\newcommand{\Chifab}{\boldsymbol{\chi}^{\text{fab}}}
\newcommand{\Chistruct}{\boldsymbol{\chi}^{\text{str}}}
\newcommand{\ChiStruct}{\boldsymbol{\chi}^{\text{Str}}}
\newcommand{\Chipre}{\boldsymbol{\chi}^{\text{pre}}}
\newcommand{\Sigmabold}{\boldsymbol{\sigma}}
\newcommand{\Chibold}{\boldsymbol{\chi}}
\newcommand{\Sigmao}{\Sigmabold_{\!\text{o}}}
\newcommand{\Sigmadot}{\dot{\Sigmabold}}
\newcommand{\Psio}{\psi_{\text{o}}}
\newcommand{\Varepsilono}{\boldsymbol{\varepsilon}_{\text{o}}}
\newcommand{\Varepsilonp}{\boldsymbol{\varepsilon}^{\prime}}
\newcommand{\Varepsilonpp}{\boldsymbol{\varepsilon}^{\prime\prime}}

\newcommand{\AlphaStructpp}{\boldsymbol{\alpha}^{\prime\prime,\text{Str}}}

\newcommand{\AlphaStructp}{\boldsymbol{\alpha}^{\prime,\text{Str}}}
\newcommand{\Alphafabp}{\boldsymbol{\alpha}^{\prime,\text{fab}}}
\newcommand{\Alphastructo}{\boldsymbol{\alpha}^{\text{str}}_{\text{o}}}
\newcommand{\AlphaStructo}{\boldsymbol{\alpha}^{\text{Str}}_{\text{o}}}
\newcommand{\Varepsilon}{\boldsymbol{\varepsilon}}
\newcommand{\Varepsilondot}{\dot{\boldsymbol{\varepsilon}}}
\newcommand{\Erev}{\mathbf{H}^{\Varepsilon\Varepsilon}}
\newcommand{\Erevn}{H^{\Varepsilon\Varepsilon}}

\newcommand{\varrhoc}{\varrho^{\text{c}}}
\newcommand{\varrhocsl}{\varrho^{\text{c,sl}}}
\newcommand{\Xibold}{\boldsymbol{\xi}}

\newcommand{\Nabla}{\boldsymbol{\nabla}}
\newcommand{\Nablaalpha}{\boldsymbol{\nabla}_{\!\Alphasett}\!}
\newcommand{\BDelta}{\boldsymbol{\delta}}
\newcommand{\Deltaslip}{\dot{\boldsymbol{\delta}}^{\,\text{t,sl}}}
\newcommand{\deltaslip}{\dot{\delta}^{\,\text{t,sl}}}
\newcommand{\Ft}{\mathbf{f}^{\,\text{t}}}
\newcommand{\Fst}{f^{\,\text{t}}}
\newcommand{\Esl}{E^{\text{c,sl}}}
\newcommand{\Ec}{E^{\text{c}}}
\newcommand{\Legendre}{\Psi}
\makeatletter
\def\ps@pprintTitle{%
  \let\@oddhead\@empty
  \let\@evenhead\@empty
}

\begin{document}
%
\begin{tcolorbox}[
    colback=gray!5, 
    colframe=gray!70, 
    arc=0mm, 
    width=\textwidth,
    boxrule=0.5pt,
    top=2mm, bottom=2mm, left=2mm, right=2mm
]
\small
\textbf{Please cite this article as:} \\
Matthew R. Kuhn (2026),
  ``Thermomechanics of dense granular materials: A particle-scale perspective,''
  \emph{Journal of the Mechanics and Physics of Solids},
  Vol. 216,
  106749,
https://doi.org/10.1016/j.jmps.2026.106749.
\end{tcolorbox}
\vspace{1em} 
\begin{frontmatter}
\title{\Large\bfseries
       Thermomechanics of dense granular materials:
       a particle-scale perspective\normalfont}
%
%
%
\author{Matthew R. Kuhn}
%
\affiliation{organization={Emeritus Professor,
                           Donald P. Shiley School of Engr.,
                           Univ. of Portland},
             city={Portland},
             state={Oregon},
             country={U.S.A.}}
\begin{abstract}\small
The paper presents a broad thermomechanic framework for the
isothermal rate-independent constitutive behavior of dense
granular materials.
The essential quantities in this framework are directly
measurable in discrete element (DEM) simulations:
free energy, dissipation, stress, and strain.
The paper proposes that energy and dissipation
are governed by two sets of internal variables:
fabric variables that control the reversible stiffness
and structure variables associated with internal sliding.
The relevant fabric variables are identified and measured with
simulations.
Two hypotheses are considered for the structure variables:
the macro-scale irreversible strain and
an aggregate measure of the micro-scale frictional forces among
sliding contacts.
Both hypotheses are tested with simulations,
which allow direct calculation of the internal variables.
The paper then demonstrates the manner in which the measured
variables are applied in incremental constitutive models.
Among other findings are the following.
(1)~Dissipation from contact sliding is pervasive and
occurs in all directions of incremental loading.
(2)~Contact motions are not reversed by a reversal of the
strain direction, and contacts continue to slide
when loading is reversed.
(3)~The free energy can not be assumed smoothly
differentiable; instead, G\^{a}teaux
derivatives must be used with irreversible effects.
(4)~Basic assumptions of elastoplasticity are contravened:
no region of purely reversible strain exists,
no uniform yield direction exists,
no uniform flow direction exists,
and irreversible strain is not proportional
to the projected total strain.
A three-mechanism elastoplasticity model, however,
closely fit the DEM results, and methods are demonstrated
for quantifying the model.
The results emphasize that advanced constitutive models are needed
for capturing the general incremental behavior
of granular materials.
\end{abstract}
\end{frontmatter}
%
\section{\normalsize Introduction}
Thermomechanics is an encompassing framework for 
understanding and modeling material behavior,
accounting for heat exchange and energy dissipation
within deformable bodies.
The field has its 20th century origins in the works
of Truesdell \cite{Truesdell:1960b,Truesdell:1960a},
Noll \cite{Noll:1958a}, and Green and Nagdhi
\cite{Green:1972a,Green:1977a},
and has been formalized in the monographs of
Ziegler \cite{Ziegler:1983a}
and Maugin \cite{Maugin:1999a}.
A fundamental concept in thermomechanics was introduced
by Coleman and Gurtin \cite{Coleman:1967a},
who proposed that a material's
internal energy depends not only on observable state variables~---
such as deformation and temperature~---
but also on a set of internal ``hidden'' variables.
These internal variables are typically associated with
irreversible processes,
including plastic and viscous deformation, dislocation mobility, and
internal damage \cite{Rice:1971a,Lubliner:1972a,Germain:1983a}.
By incorporating a dissipation function that depends on these variables,
Ziegler and Maugin systematically
developed thermomechanic models of
elastoplasticity and viscoplasticity
\cite{Ziegler:1983a,Maugin:1999a}.
\par
The application of thermomechanics to geotechnical materials,
including granular materials, was pioneered by Collins and
colleagues \cite{Collins:1997a,Collins:2002a,Collins:2005a},
who established a
consistent framework for rate-independent
and pressure-dependent materials under isothermal conditions.
Among their accomplishments,
they formalized the distinction between elastic and
irreversible strains when elastic moduli are coupled with
internal variables \cite{Collins:1997a}.
They also derived elastoplastic models for soils and
introduced the concept of stored plastic work to explain
dilatancy and the effects of over-consolidation \cite{Collins:2005b}.
\par
Thermomechanic methods have since been applied to a range of problems,
including thermally induced deformation in sands \cite{Pan:2023a},
the separate treatment of mechanical and hydraulic free energy
in unsaturated soils \cite{Zhao:2019a},
heat conduction in granular media \cite{Nguyen:2009a},
and heat generation in granular flows \cite{Rangel:2024a}. 
\par
Building on Collins’ work,
the paper examines the thermomechanics of granular materials,
making use of particle-level discrete element (DEM) simulations
to directly quantify and examine
the notions of internal energy and dissipation.
The simulated system consists of an unbonded assembly
of durable (non-breaking) particles,
with contact interactions idealized as springs
in series with frictional sliders.
Such simulations, now routinely used in geomaterials modeling,
allow measurement of thermomechanic
quantities that are not accessible in conventional
laboratory experiments.
Using these measurements,
we appraise fundamental assumptions of elastoplasticity,
the dominant constitutive approach for
soils and other granular materials.
Although elastoplasticity has been refined extensively
to better match observed behavior~---
non-smooth yield and plastic potentials,
multi-modal yield mechanisms,
bounding surfaces,
uncoupled stiffness,
multiple constitutive cones, etc.
(one is reminded of Ptolemy's epicycles)~---
we return to fundamental thermomechanic principles
to reassess granular behavior.
The simulation results reveal unexpected behaviors that
contravene many conventional assumptions of elastoplasticity
and that highlight misunderstandings of granular behavior
and the limitations of current models.
\par
To accommodate these results,
we propose a broad class of energy functions that are less smooth than
those typically assumed in elastoplastic models.
We also identify a set of internal variables
particularly relevant to granular materials,
including contact density, contact anisotropy,
sliding density, and sliding stress.
Our purpose, however, is not to develop a detailed constitutive model.
Instead, we present two constitutive frameworks~--- macro and micro~---
and demonstrate how their key elements can be
measured with simulations.
\par
The paper begins by recounting general
thermomechanic principles of rate-independent
materials under isothermal conditions,
and by introducing a set of possible constitutive forms
(Section~\ref{sec:principles}).
We then propose two sets of internal variables
relevant to granular materials:
fabric variables, which govern the reversible moduli,
and structure variables, which control dissipation
(Section~\ref{sec:viewpoints}).
The structure variables can be defined in two ways.
In a macro-scale approach, consistently with elastoplasticity,
the structure variables are taken as the irreversible strain.
Alternatively,
a micro-scale approach defines the structure variables 
using directly measured micro-scale quantities:
the aforementioned sliding density and sliding stress.
Section~\ref{sec:dem} describes the DEM simulations and
identifies the fabric variables that determine
the elastic moduli and drive the (coupling) dissipation
associated with their evolution.
Sections~\ref{sec:macroresults} and~\ref{sec:microresults}
use the simulations to evaluate the
macro-scale (elastoplastic) and micro-scale (sliding-based)
constitutive frameworks, assess their validity, and
guide the development of improved models.
The closure summarizes the main findings
and addresses issues raised by the results.
\section{\normalsize Thermomechanic preliminaries}\label{sec:principles}
\par
We begin with the scalar Helmholtz free energy density
$\psi$ per unit of volume,
representing a material's latent capacity 
to supply energy to its surroundings.
For the isothermal conditions considered herein, $\psi$
is a function of strain $\boldsymbol{\varepsilon}$,
and owing to the material's inelasticity,
$\psi$ also depends on other factors,
which are taken as
a set of internal variables
$\Alphaset$,
containing scalars, vectors,
and/or tensors as its elements
\cite{Rice:1971a,Lubliner:1972a,Maugin:1994a}.
This set $\Alphaset$ includes variables that quantify
a granular material's fabric-related particle arrangement and
its structure-related inter-particle forces.
\par
We assume that the material is rate-independent, and
because isothermal conditions are assumed,
the Helmholtz energy is a function of form
$\psi = \psi(\Varepsilon,\Alphaset)$.
For general three-dimensional conditions and
a total of $m$ internal variables $\alpha_{i}$
in $\Alphaset$,
the domain of $\psi$ is the set of states 
$\Omega\subset\mathbb{R}^{6}\times\mathbb{R}^{m}$,
with $\psi\colon \Omega\rightarrow\mathbb{R}$;
whereas
for the simpler conventional triaxial conditions considered
herein, having only two independent generalized strains,
$\Omega\subset\mathbb{R}^{2}\times\mathbb{R}^{m}$,
with $\psi\colon \Omega\rightarrow\mathbb{R}$.
We assume that function $\psi$ is continuous and
is twice continuously differentiable
with respect to strain $\Varepsilon$.
However,
less restrictive
conditions are assumed for set $\Alphaset$:
although $\psi$ is assumed a continuous function
of both $\Varepsilon$ and $\Alphaset$, we allow
the possibility that $\psi$ is not continuously differentiable
with respect to $\Alphaset$.
In particular, we encounter evidence that the energy rate $\dot{\psi}$
\emph{depends on the direction of} $\Alphadotset$
and thus adopt broader assumptions that are still
consistent with the rate-independence of $\psi$.
\par
This question of smoothness requires attention to
notation.
For a scalar function $f$ of a vector argument $\mathbf{z}$,
the standard gradient
$\Nabla_{\!\mathbf{z}}f=\partial f/\partial\mathbf{z}$
applies when $f$ is differentiable with respect to $\mathbf{z}$.
In this case, rate $\dot{f}$ is the inner product of
gradient vector
$\Nabla_{\mathbf{z}}f$
and rate $\dot{\mathbf{z}}$, or
$\dot{f}=\Nabla_{\mathbf{z}}f\cdot\dot{\mathbf{z}}$.
The more general
G\^{a}teaux derivative applies when
$f$ is not differentiable but is directionally differentiable.
In this case, the derivative of $f$
depends on the direction of rate $\dot{\mathbf{z}}$,
but not by way of a gradient vector.
We use the following notation for G\^{a}teaux derivatives:
\begin{equation}\label{eq:Gateaux}
  \Nabla_{\mathbf{z}}[\Xibold]f
  =
  \frac{\partial_{\Xibold}f(\mathbf{z})}
        {\partial\mathbf{z}}
  =
  \left(
  \lim\limits_{h \to 0^{+}}
  \frac{f(\boldsymbol{x} + h\Xibold)
        - f(\boldsymbol{x})}
       {h}
   \right)
  \Xibold
\end{equation}
in which
scalar $h\in\mathbb{R}^{+}$ approaches $0$ from the right,
and $\Xibold$ is the direction of change of
$\mathbf{z}$
(the direction of $\dot{\mathbf{z}}$),
or $\Xibold=\dot{\mathbf{z}}/\|\dot{\mathbf{z}}\|$,
where $\|\cdot\|$ is the norm.
Multiplying the limit by
direction $\Xibold$ in Eq.~(\ref{eq:Gateaux}) renders the rate $\dot{f}$
as an inner product, such that
$\dot{f}=\Nabla_{\mathbf{z}}[\Xibold]f\cdot\dot{\mathbf{z}}$,
but one that explicitly depends on direction $\Xibold$.
\par
With this understanding,
force quantities are
defined as derivatives of the Helmholtz energy,
\begin{equation}\label{eq:psiderivs}
\begin{alignedat}{3}
  &\Sigmabold\big(\Varepsilon,\Alphaset\big) &&=
  \Nabla_{\Varepsilon}\psi &&=
  \frac{\partial\psi\big(\Varepsilon,\Alphaset\big)}
       {\partial\Varepsilon}
  \\
  &\left\{\boldsymbol{\chi}\big(\Varepsilon,\Alphaset;\Xibold\big)\right\} &&=
  -\Nablaalpha[\Xibold]\psi &&=
  -\left(
  \frac{\partial_{\boldsymbol{\xi}}\psi\big(\Varepsilon,\Alphaset\big)}
        {\partial\Alphaset}
  \right)
  \Xibold
\end{alignedat}
\end{equation}
where the true gradient $\Sigmabold$ is the conventional stress
(herein, the Cauchy stress),
and the G\^{a}teaux
derivative $\Chiset$ is a set of dissipative
stress-like quantities that are complementary to the
internal variables $\Alphaset$.
Henceforth, $\Xibold$ is taken
as the unit direction of rate $\Alphadotset$,
\begin{equation}
  \Xibold =
  \Alphadotset \,/\, \|\Alphadotset\|
\end{equation}
where $\|\cdot\|$ is an appropriate norm.
\par
Energy $\psi$ is usually assumed continuously
differentiable with respect to $\Alphaset$,
allowing the simpler, conventional gradient
$\Chiset=\Nabla_{\boldsymbol{\alpha}}\psi
=\partial\psi/\partial\Alphaset$.
Experiments herein demonstrate, however,
that derivative $\Chiset$ can change abruptly upon
a change in the direction of rate $\Alphadotset$,
a result that is inconsistent with a continuously differentiable
$\psi$.
Although the projection of $\psi$ onto
the $\Varepsilon$-space is differentiable,
its projection onto $\Alphaset$ reveals sharper ridges
and valleys, making necessary the more general G\^{a}teaux
derivative.
\par
The energy rate
\begin{equation}\label{eq:balance}
  \dot{\psi}(\Varepsilon,\Alphaset) =
  \Sigmabold\cdot \Varepsilondot
  -
  \Chiset\cdot \Alphadotset
\end{equation}
expresses the material's energy balance, 
in which product
$\Sigmabold\cdot \dot{\Varepsilon}$
is the rate of stress-work,
and product
$\Chiset\cdot \Alphadotset$
is the dissipation rate
(i.e., internal entropy production),
the latter product denoting the sum of individual
inner products,
$\Chiset\cdot \Alphadotset=\sum\boldsymbol{\chi}_{i}\cdot\dot{\boldsymbol{\alpha}}_{i}$.
The second law stipulates that
dissipation is non-negative,
\begin{equation}\label{eq:ineq}
  \Chiset\cdot \Alphadotset \ge 0
\end{equation}
noting that the inequality applies to the aggregate set
$\Alphadotset$, 
although we encounter instances in which individual products
$\boldsymbol{\chi}_{i}\cdot\dot{\boldsymbol{\alpha}}_{i}$ are negative.
\par
Rates of the stress quantities
follow from Eq.~(\ref{eq:psiderivs}):
\begin{align}\label{eq:sigmadot}
  \dot{\Sigmabold} &=
  \Erev(\Varepsilon,\Alphaset)
  \cdot
  \dot{\Varepsilon}
  \;+\:
  \Nablaalpha
  [\Xibold]\boldsymbol{\sigma}
  \cdot
  \Alphadotset
  ,\quad\;
  \Erev(\Varepsilon,\Alphaset)
  =
  \Nabla_{\Varepsilon}\Sigmabold
  =
  \frac{\partial^{2}\psi}
       {\partial\Varepsilon\partial\Varepsilon}
  \\
  \label{eq:chidot}
  \left\{\dot{\boldsymbol{\chi}}\right\} &=
  -\Nabla_{\Varepsilon}\Chiset
  \cdot
  \dot{\Varepsilon}
  \;-\:
  \Nablaalpha
  [\Xibold]\Chiset
  \cdot
  \Alphadotset
\end{align}
In the first equation, we have assumed that $\psi$
is twice differentiable with respect to strain $\Varepsilon$
(an assumption supported by evidence herein and elsewhere),
so that rate \mbox{$\Erev\cdot\Varepsilondot$}
is a linear transformation
of $\Varepsilondot$.
Fourth-order tensor $\Erev$
is the reversible stiffness modulus~---
the material's stiffness when dissipative changes in
$\boldsymbol{\alpha}$ are prevented
(e.g., by disallowing frictional slips between particles).
Because of the assumed dependence of
$\Erev$ on internal variables,
the material is considered a \emph{coupled material},
a characteristic of materials in
which the elastic moduli
depend on the plastic strain,
with
$\partial^{3}\psi/\partial\Varepsilon\partial\Varepsilon\partial\Alphaset \neq 0$.
\par
In Eq.~(\ref{eq:sigmadot}),
we apply the more general derivative
$\Nablaalpha[\boldsymbol{\xi}]\Sigmabold$
of Eq.~(\ref{eq:Gateaux}),
allowing possible direction-dependence of stress
$\Sigmabold$ on rate $\Alphadotset$.
With Eq.~(\ref{eq:chidot}),
we have assumed that set $\Chiset$ is differentiable
with respect to $\Varepsilon$, so that
$\Nabla_{\Varepsilon}\Chiset$
is the true gradient of
$\Chiset$ with respect to $\Varepsilon$.
The second term in Eq.~(\ref{eq:chidot}), however, assumes
a direction-dependent G\^{a}teaux derivative of
$\Chiset$ with respect to $\Alphaset$.
\par
The principles expounded above
apply in a continuum setting,
in which the essential quantities~---
stress, strain, energy density, etc.~---
are defined at idealized
material points within a continuous body.
When granular materials and other discontinua
are treated as continua,
these quantities are typically interpreted over representative
volume elements (RVEs), for which measurement inevitably
requires spatial averaging within the RVE combined
with a temporal smoothing across time.
\par
The RVE considered herein is a numerical
assembly of discrete non-spherical
particles that is analyzed with discrete element (DEM) simulations
(Section~\ref{sec:dem}).
These simulations, in its essence, treat a granular assembly as
a system of nodes connected with springs and sliders (in series),
but a system whose connection topology is continually
rearranged by the breaking and forming of contacts
during bulk deformation.
With particle-scale simulations,
many of the essential thermomechanic quantities
can be directly measured as aggregates of the springs and sliders,
and such measurements
can be obtained at any equilibrium state along a simulation's
loading path (i.e., thermomechanic process).
These measurable quantities are
listed in Table~\ref{table:measure} and
their spatial averaging is described in \ref{sec:measuredem}.
Besides measuring current values,
a series of
DEM probes can also be conducted,
starting from any equilibrium state,
to investigate a quantity's directional rate
with respect to the prescribed strain directions,
$\Varepsilondot$.
It should be noted, however, that
the internal variables
$\Alphaset$
and their stress-like complements $\Chiset$
can not be fully controlled or measured,
even with DEM simulations \cite{Bridgman:1953a};
consequently, neither set is included in Table~\ref{table:measure}.
At best,
$\Alphaset$ and $\Chiset$ can be inferred if
one adopts sufficient discernment
of their character,
as is attempted in Section~\ref{sec:viewpoints}.
The DEM simulations are then used to test this reckoning.
\begin{table}
  \centering\small
  \caption{\small Quantities directly measured in
           simulations.  The upper rows are
           quantities measurable at any
           equilibrium state (see \ref{sec:measuredem}).
           Combinations of the stress and strain rates,
           $\Sigmadot$ and
           $\Varepsilondot$, can also be controlled in DEM simulations.
           The bottom two rows are quantities computed along
           a path, provided that modulus
           $\Erev$ is first determined.
           \label{table:measure}}
  \begin{tabular}{cp{9cm}}
    \toprule
    Quantity & Methods\\
    \midrule
    $\Sigmabold$, $\Sigmadot$
      &
      stress computed with
      Eq.~(\ref{eq:stress}).
      \\
    $\Varepsilon$, $\Varepsilondot$
      &
      strain measured at
      assembly's boundaries.
      \\
    $\psi$ & volume average of elastic energy, Eq.~(\ref{eq:Helm}).
      \\
    $\Chiset\cdot \Alphadotset$
      &
      full dissipation rate, Eq.~(\ref{eq:DEMdissip})
      \\
    $\Erev$
     &
     reversible stiffness moduli,
$\mathbf{E}^{\text{rev}}=\partial\Sigmabold/\partial\Varepsilon
     |_{\boldsymbol{\alpha}=\text{const}}=
     \partial^{2}\psi / \partial\Varepsilon^{2}$,
     measured by ``locking'' the contacts by preventing their
     frictional slip, as in \cite{Kuhn:2018a,Kuhn:2018c}.
     \\
     \midrule
     $\mathbf{F}_{1}$ &
     cumulative effect of variables $\Alphaset$ on stress,
     Eq.~(\ref{eq:sigmagen})
     \\
     $F_{2}$ &
     cumulative dissipative effect of $\Alphaset$ on energy $\psi$,
     Eq.~(\ref{eq:psigen})
     \\
  \bottomrule
  \end{tabular}
\end{table}
\par
Before addressing these matters,
we consider the general constitutive form of
Helmholtz energy $\psi$,
a form that we later quantify with simulations.
We start with
the reversible stiffness modulus $\Erev=
\partial\Sigmabold/\partial\Varepsilon$,
which appears in the stress-rate of
Eq.~(\ref{eq:sigmadot}) and
can be fully measured in all strain directions
$\Varepsilondot$ at points
along a loading path
(Section~\ref{sec:moduli} and \ref{sec:measuredem}).
Once $\Erev$ is measured,
Eqs.~(\ref{eq:psiderivs}\textsubscript{1})
and~(\ref{eq:sigmadot}\textsubscript{2}) can be integrated,
beginning from the initial state
$(\Varepsilono,\Alphaset_{\text{o}})$,
yielding the following general
forms of stress and energy as line integrals
along a $(\Varepsilon,\Alphaset)$ path:
\begin{align}
  \label{eq:sigmagen}
  &\Sigmabold(\Varepsilon,\Alphaset)
  =
  \int_{\Varepsilono}^{\Varepsilon}\!\!
    \Erev
    (\Varepsilon^{\prime},
     \{\boldsymbol{\alpha}^{\prime}\})
  \cdot
  d\Varepsilon^{\prime}
  \:+\:
  \mathbf{F}_{1}(\Alphaset)
  \:+\:
  \Sigmao
  \\
  \label{eq:psigen}
  &\begin{aligned}
  \psi(\Varepsilon,\Alphaset)
  =&
  \int_{\Varepsilono}^{\Varepsilon}\!
  d\Varepsilon^{\prime\prime}
  \cdot\!  \int_{\Varepsilono}^{\Varepsilonpp}\!\!\!
  \Erev
    (\Varepsilon^{\prime},\{\boldsymbol{\alpha}^{\prime}\})
  \cdot
  d\Varepsilon^{\prime}
  \\
  &\:+\:
  \int_{\Varepsilono}^{\Varepsilon}\!
  \mathbf{F}_{1}(\{\boldsymbol{\alpha}^{\prime}\})
  \cdot d\Varepsilon^{\prime}
  \:+\:
  F_{2}(\Alphaset)
  \;+\;
  \Sigmao\!\cdot(\Varepsilon-\Varepsilono)
  \:+\:
  \Psio
  \end{aligned}
\end{align}
in which
$\Sigmao$, $\Psio$, and $\Varepsilono$
are initial values;
$\mathbf{F}_{1}$ is a stress-like tensor function
of set $\Alphaset$ that accounts
for cumulative irreversible stress effects; and
$F_{2}$ is a scalar function of $\Alphaset$ associated
with cumulative dissipation.
Other than
the two functions
$\mathbf{F}_{1}$ and $F_{2}$,
all quantities in Eqs.~(\ref{eq:sigmagen})
and~(\ref{eq:psigen})
are measurable in DEM simulations.
Therefore, $\mathbf{F}_{1}$ and $F_{2}$
can be separately computed at all points on a loading path:
one first computes
$\mathbf{F}_{1}(\Alphaset)$ by applying Eq.~(\ref{eq:sigmagen});
one then computes $F_{2}(\Alphaset)$ with Eq.~(\ref{eq:psigen}).
Because they are directly computed from data along a path,
$\mathbf{F}_{1}$ and $F_{2}$ are included in Table~\ref{table:measure}.
This evaluation of $\mathbf{F}_{1}$ and $F_{2}$ can be done
\emph{without regard to the natures or values of the internal variables
$\Alphaset$ that affect the two functions}.
The two functions will take a central role in
quantifying and appraising the chosen sets $\Alphaset$.
\par
The free energy $\psi$ in Eq.~(\ref{eq:psigen})
is computed at points along a path in the state space of
$\Varepsilon$ and $\Alphaset$ between the initial and final
states, say $(\Varepsilono,\{\boldsymbol{\alpha}_{\text{o}}\})$
and $(\Varepsilon_{\text{f}},\{\boldsymbol{\alpha}_{\text{f}}\})$.
With a proper choice of internal variables $\Alphaset$,
the final energy $\psi_{\text{f}}$ at
 $(\Varepsilon_{\text{f}},\{\boldsymbol{\alpha}_{\text{f}}\})$
should be independent of the path traversed between the two states.
The gradient theorem holds that a
sufficient condition for path-independence is the
differentiability of $\psi$ with respect to
$\Alphaset$ along all such paths.
Two sets of internal variables are proposed in 
Section~\ref{sec:viewpoints},
differentiability with each set is measured in
Sections~\ref{sec:dem}--\ref{sec:microresults},
and the path-independence of $\psi$ is assessed in 
Section~\ref{sec:conclusions}.
\subsection{Auxiliary conditions and constitutive forms}%
\label{sec:constituteforms}
A material's constitutive description is a relationship
among the stress and strain rates,
$\Sigmadot$ and $\Varepsilondot$, for a given state
$(\varepsilon,\Alphaset)$,
allowing solution of one set of rates (the response)
from knowledge of the other rates (the loading).
A thermomechanic framework provides the means
of developing constitutive relations and, conversely,
of determining whether a given constitutive model
comports with a material's behavior.
Three incremental forms of a material's constitutive are now described,
as these forms are the context for adopting
DEM results within continuum models.
In the first incremental form,
the rates $\Sigmadot$ and $\Chidotset$, which are
derived from energy
$\psi(\Varepsilon,\Alphaset)$
in Eqs.~(\ref{eq:sigmadot})--(\ref{eq:chidot}), are expressed
in matrix form as
\begin{gather}\label{eq:matrix1}
 \text{Form~1:}\quad
 \left[
   \begin{array}{@{\extracolsep{0pt}}c@{\extracolsep{0pt}}}
     \Sigmadot
     \\\hdashline[1pt/1pt]
     \Chidotset
   \end{array}
 \right]_{(6+m)\times 1}
 \!=\;
 \left[
   \begin{array}%
         {@{\extracolsep{0ex}}c;{1pt/1pt}c@{\extracolsep{\fill}}}
     \Erev & \mathbf{H}^{\Varepsilon\boldsymbol{\alpha}}(\Xibold)
     \\\hdashline[1pt/1pt]
     \mathbf{H}^{\boldsymbol{\alpha}\Varepsilon}\rule{0pt}{2.2ex}
     &
     \mathbf{H}^{\boldsymbol{\alpha}\boldsymbol{\alpha}}(\Xibold)
   \end{array}
 \right]
 \left[
    \begin{array}{@{\extracolsep{0pt}}c@{\extracolsep{0pt}}}
        \Varepsilondot
        \\\hdashline[1pt/1pt]
        \Alphadotset
    \end{array}
  \right]_{(6+m)\times 1}
  \\
  \mathbf{H}^{\Varepsilon\boldsymbol{\alpha}}(\Xibold)
  = \Nablaalpha [\Xibold]\Sigmabold
  ,\quad
  \mathbf{H}^{\boldsymbol{\alpha}\Varepsilon}
  = -\Nabla_{\Varepsilon}\Chiset
  ,\quad
  \mathbf{H}^{\boldsymbol{\alpha}\boldsymbol{\alpha}}(\Xibold)
  = -\Nablaalpha [\Xibold]\Chiset
\end{gather}
where $m$ is the number of components in $\Alphaset$;
$[\Sigmadot]$ and $[\Varepsilondot]$ are
$6\times1$ vectors of their tensor components;
$\Chidotset$ and $\Alphadotset$ are $m\times 1$ vectors;
$\Erev$ is the $6\times6$ matrix of its
components in
Eq.~(\ref{eq:sigmadot}\textsubscript{2});
$\mathbf{H}^{\Varepsilon\boldsymbol{\alpha}}$
is an $6\times m$ matrix;
$\mathbf{H}^{\boldsymbol{\alpha}\Varepsilon}$ is an $m\times 6$ matrix;
and
$\mathbf{H}^{\boldsymbol{\alpha}\boldsymbol{\alpha}}$ is an $m\times m$
matrix.
The elements of all four sub-matrices in Eq.~(\ref{eq:matrix1})
are functions of the state $(\Varepsilon,\Alphaset)$.
Note, if $\psi$ is not differentiable with respect
to $\Alphaset$,
Schwarz's theorem does not apply, so that
the off-diagonal matrices,
$\mathbf{H}^{\Varepsilon\boldsymbol{\alpha}}$ and
$\mathbf{H}^{\boldsymbol{\alpha}\Varepsilon}$, are not
necessarily the transpose of each other.
Finally, the size of~6 in the equations is reduced to~2
for simplified, triaxial conditions (see Eq.~\ref{eq:components}).
\par
Unless the material behaves elastically,
Eq.~(\ref{eq:matrix1}) alone
is insufficient
to establish a unique relationship between the
rates of stress and strain, $\Sigmadot$ and $\Varepsilondot$:
knowing $\Varepsilondot$ does not
yield a unique solution for $\Sigmadot$,
when the rates $\Chidotset$ are also unknown.
Additional information is required
to establish this relationship.
\par
Another means of expressing a material's
constitutive behavior is to introduce
a set $\{\boldsymbol{\phi}\}$ of $n$
auxiliary constraints $\phi_{i}$ of the form
\begin{equation}\label{eq:constr}
  \phi_{i}(\Varepsilon,\Alphaset,\Varepsilondot,\Alphadotset)=0,
  \quad
  i = 1,2,\ldots,n
\end{equation}
One can view these equations as internal constraints on
the variables $\Alphaset$,
whose rates are coerced by the controlling strain $\Varepsilondot$.
Because the material is assumed rate-independent,
the constraint functions must be positive-homogeneous
of degree~1 in the rates $\Varepsilondot$
and $\Alphadotset$,
such that
$\phi_{i}(\Varepsilon,\Alphaset,\lambda\Varepsilondot,\lambda\Alphadotset)
=\lambda\phi_{i}(\Varepsilon,\Alphaset,\Varepsilondot,\Alphadotset)=0$,
$\lambda\in\mathbb{R}$.
Each function $\phi_{i}$, therefore, must satisfy the Euler condition
\begin{equation}\label{eq:euler}
  \phi_{i}\left(\Varepsilon,\Alphaset,\Varepsilondot,\Alphadotset\right)
  \;=\;
  \frac{\partial\phi_{i}}{\partial\Varepsilondot}
    \cdot
    \Varepsilondot
  \;+
  \sum_{\dot{\alpha}_{j}
        \in\Alphadotset}
  \frac{\partial\phi_{i}}{\partial\dot{\alpha}_{j}}
    \cdot
    \dot{\alpha}_{j}
  \;=\;
  0
\end{equation}
Although a material can, in principle,
reach the full range of strain directions $\Varepsilondot$,
the range of the directions $\boldsymbol{\xi}$ of
rates $\Alphadotset$ are limited by the
constraints $\{\boldsymbol{\phi}\}$.
For example, if $\boldsymbol{\alpha}$ is taken as
the irreversible strain in a conventional elastoplastic model,
the response $\dot{\boldsymbol{\alpha}}$
is confined to the single direction defined by the plastic flow rule.
\par
When the constraints $\{\boldsymbol{\phi}\}$ are known and are
of a sufficient number $n$, Eq.~(\ref{eq:matrix1}) can be replaced
by the following incremental equations:
\begin{gather}\label{eq:matrix4}
 \text{Form~2:}\quad
 \left[
   \begin{array}{@{\extracolsep{0pt}}c@{\extracolsep{0pt}}}
     \Sigmadot
     \\\hdashline[1pt/1pt]
     \mathbf{0}
   \end{array}
 \right]_{(6+n)\times 1}
 \!=\;
 \left[
   \begin{array}%
         {@{\extracolsep{0ex}}c;{1pt/1pt}c@{\extracolsep{\fill}}}
     \Erev & \mathbf{H}^{\Varepsilon\boldsymbol{\alpha}}(\Xibold)
     \\\hdashline[1pt/1pt]
     \mathbf{C}^{\Varepsilon}\rule{0pt}{2.2ex}
     &
     \mathbf{C}^{\boldsymbol{\alpha}}(\Xibold)
   \end{array}
 \right]
 \left[
    \begin{array}{@{\extracolsep{0pt}}c@{\extracolsep{0pt}}}
        \Varepsilondot
        \\\hdashline[1pt/1pt]
        \Alphadotset
    \end{array}
  \right]_{(6+m)\times 1}
  \\
  \label{eq:defHea}
  \mathbf{C}^{\Varepsilon}
  = \Nabla_{\Varepsilon} \{\boldsymbol{\phi}\}
  ,\quad
  \mathbf{C}^{\boldsymbol{\alpha}}(\Xibold)
  = \Nabla_{\boldsymbol{\alpha}}[\Xibold]\{\boldsymbol{\phi}\}
\end{gather}
in which $\mathbf{C}^{\Varepsilon}$
and $\mathbf{C}^{\boldsymbol{\alpha}}$
are $n\times 6$ and $n\times m$ matrices.
The lower set of equations in~(\ref{eq:matrix4}) are non-linear:
although the product
$\mathbf{C}^{\Varepsilon}\cdot\Varepsilondot$
is assumed linear in $\Varepsilondot$, and
$\mathbf{C}^{\boldsymbol{\alpha}}(\Xibold)\cdot\Alphadotset$
is homogeneous in $\Alphadotset$, the latter products
will likely depend upon the directions $\Xibold$
of rates $\Alphadotset$.
\par
The constraints must satisfy
the second law, requiring the dissipation rate
$\Chiset\cdot\Alphadotset$ to be non-negative,
so that half of the $\mathbb{R}^{m}$ hyper-space
of rates $\Alphadotset$ is disallowed (see Eq.~\ref{eq:ineq}).
Thus, to heed the second law, one must monitor
the set $\Chiset$ by advancing
this set with the rates $\Chidotset$ computed
with Eq.~(\ref{eq:matrix1}).
\par
A third constitutive approach is that of
elastoplastic models, which emerge from
a dissipation function $\varphi$
that provides the dissipation rate $\Chiset\cdot\Alphadotset$,
with $\varphi$ being a function of
the current state, $\Varepsilon$ and $\Alphaset$,
and of the rates $\Alphadotset$:
\begin{equation}\label{eq:D}
  \Chiset\cdot \Alphadotset
 \;=\;
  \varphi\big(\Varepsilon,\Alphaset,\Alphadotset\big)
\end{equation}
as in \cite{Ziegler:1983a,Ziegler:1987a,Collins:1997a}.
This dissipation
function $\varphi$,
from which the yield condition is obtained,
provides the additional information that allows
elimination of $\Chidotset$ from Eq.~(\ref{eq:matrix1}).
Because the material is assumed rate-independent,
function $\varphi$ does not depend on the strain rate
$\Varepsilondot$, as would apply
with visco-plastic materials.
Setting a constraint function $\phi_{0}$ equal to the difference
between the right and left sides of Eq.~(\ref{eq:D}),
and applying the Euler theorem~(\ref{eq:euler}), gives
\begin{equation}\label{eq:orthog1}
  \phi_{0}
    \left(\Varepsilon,\Alphaset,\Varepsilondot,\Alphadotset\right)
  \;\equiv\;
  \left(
  \frac{\partial\varphi}{\partial\Alphadotset}
  \;-\;
  \Chiset
  \right)
  \cdot
  \Alphadotset
  \;=\;
  0
\end{equation}
This constraint states that the difference within
parentheses must be orthogonal to the rate $\Alphadotset$.
\par
Ziegler \cite{Ziegler:1983a,Ziegler:1987a} proposed
that the derivatives $\partial\varphi/\partial\Alphadotset$
in Eq.~(\ref{eq:orthog1}) are \emph{individually} equal to the
dissipative stresses $\Chiset$, leading to
the stronger orthogonality condition
\begin{equation}\label{eq:orthog2}
  \frac{\partial\varphi}{\partial\Alphadotset}
  \;=\;
  \Chiset
\end{equation}
This relation asserts a component-wise equivalence of quantities on
the left and right sides of the equation.
The \emph{Ziegler orthogonality principle}, as
expressed with Eq.~(\ref{eq:orthog2}), refers
to the presumed orthogonality between vector $\Chiset$
and the iso-surfaces of $\varphi(\Alphadotset)$
at the current $\Varepsilon$ and $\Alphaset$,
thus maximizing the rate of entropy production (i.e., dissipation).
Alternatively,
Eq.~(\ref{eq:orthog2}) derives from Eq.~(\ref{eq:orthog1})
if one assumes that the individual rates
in set $\Alphadotset$ are independent of each other,
meaning that any single rate $\dot{\boldsymbol{\alpha}}_{i}$
can be adjusted independently of the remaining rates
$\Alphadotset\backslash\dot{\boldsymbol{\alpha}}_{i}$
(for example, by exerting an opportune
strain $\Varepsilondot$).
If this independence holds, then
Eq.~(\ref{eq:orthog2}) applies, and
the rates of
the derivatives on the left of the equation
are equal to
the rates $\Chidotset$ in Eq.~(\ref{eq:chidot}), such that
\begin{equation}\label{eq:consist}
  \Chidotset
  \;=\;
  \frac{\partial^{2}\varphi}
  {\partial\Alphadotset\partial\Varepsilon}
  \cdot
  \Varepsilondot
  \;+\;
  \frac{\partial^{2}_{\Xibold}\varphi}
  {\partial\Alphadotset\partial\Alphaset}
  \cdot
  \Alphadotset
  =-\Nabla_{\Varepsilon}\Chiset
  \cdot
  \dot{\Varepsilon}
  -
  \Nablaalpha[\Xibold]\Chiset
  \cdot
  \Alphadotset
\end{equation}
thus assuring consistency of functions
$\psi$ and $\varphi$,
and maintaining the yield condition during continued loading
(see \cite{Collins:1997a}).
In this equation,
we allow the possibility
that the derivatives of the dissipation function $\varphi$
with respect to $\Alphaset$ depend upon direction $\Xibold$.
\par
The new information provided by dissipation function $\varphi$
in Eqs.~(\ref{eq:D})--(\ref{eq:consist}) supplements
the matrix equation (\ref{eq:matrix1}), giving a third incremental
constitutive form (see~\cite{Collins:1997a}):
\begin{gather}\label{eq:matrix3}
 \text{Form~3:}\quad
 \left[
 \begin{array}{@{\extracolsep{0pt}}c@{\extracolsep{0pt}}}
        \Sigmadot
        \\\hdashline[1pt/1pt]
        \mathbf{0}
    \end{array}
 \right]_{(6+m)\times 1}
 \!\!\!=\;
 \left[
   \begin{array}%
         {@{\extracolsep{0pt}}c;{1pt/1pt}c@{\extracolsep{\fill}}}
     \,\Erev & \mathbf{H}^{\Varepsilon\boldsymbol{\alpha}}(\Xibold)
     \\\hdashline[1pt/1pt]
     \rule{0ex}{2.3ex}
     \mathbf{H}^{\boldsymbol{\alpha}\Varepsilon}
     - \mathbf{K}^{\boldsymbol{\alpha}\Varepsilon}(\Xibold)
     &
  \mathbf{H}^{\boldsymbol{\alpha}\boldsymbol{\alpha}}(\Xibold)
  - \mathbf{K}^{\boldsymbol{\alpha}\boldsymbol{\alpha}}(\Xibold)
   \end{array}
 \right]
 \left[
    \begin{array}{@{\extracolsep{0pt}}c@{\extracolsep{0pt}}}
        \Varepsilondot
        \\\hdashline[1pt/1pt]
        \Alphadotset
    \end{array}
  \right]_{(6+m)\times 1}
  \\
  \mathbf{K}^{\boldsymbol{\alpha}\Varepsilon}(\Xibold)
  =
  \left.\partial^{2}_{\Xibold}\varphi\right/
  \partial\Alphadotset\partial\Varepsilon
  ,\quad
  \mathbf{K}^{\boldsymbol{\alpha}\boldsymbol{\alpha}}(\Xibold)
  =
  \left.\partial^{2}_{\Xibold}\varphi\right/ 
  \partial\Alphadotset\partial\Alphaset
\end{gather}
where $\mathbf{K}^{\boldsymbol{\alpha}\Varepsilon}$ and
$\mathbf{K}^{\boldsymbol{\alpha}\boldsymbol{\alpha}}$ are
$m\times6$ and $m\times m$
matrices.
In the lower part of the matrix equation~(\ref{eq:matrix3}),
the rates $\Chidotset$ have been eliminated
from Eq.~(\ref{eq:matrix1}), and
in their place
one recognizes a set $\{\boldsymbol{\phi}\}$
of $m$ constraint equations
$\phi_{i}=0$, as in Eq.~(\ref{eq:constr}),
that are generated by the Helmholtz
and dissipation functions, $\psi$ and $\varphi$,
and the assumed Ziegler condition.
One solves equation for $\Varepsilondot$
and $\Alphadotset$,
which are then used in solving $\Sigmadot$
and $\Chidotset$ with Eq.~(\ref{eq:matrix1}).
\section{\normalsize Internal variables: macro and micro viewpoints}%
\label{sec:viewpoints}
Having reviewed thermomechanic principles,
we now apply them to dense (jammed) granular materials and
use evidence gained from DEM simulations
to arrive at specific forms of energy $\psi(\Varepsilon,\Alphaset)$
in granular materials.
We separate the set $\Alphaset$
into two subsets of internal variables:
\begin{enumerate}
\item
a set of \emph{fabric} variables $\{\Alphafab\}$
that convey information about the particles' geometric arrangement,
and
\item
a set of \emph{structure} variables $\{\Alphastruct\}$
that pertain to the contact forces among particles
(this fabric--structure distinction is that of
Mitchell and Soga \cite{Mitchell:2005a}).
Specifically, the forces are of those contacts that have reached
the friction limit.
\end{enumerate}
In Section~\ref{sec:moduli2},
we demonstrate that modulus $\Erev$ depends,
almost exclusively, on strain
$\Varepsilon$ and on a small number
of fabric variables $\{\Alphafab\}$.
When the particles'
contacts are simplified with a particular linear-frictional
contact model~-- as with our simulations~---
the set $\{\Alphafab\}$ collapses to a single tensor:
the product of a scalar contact density
and a normalized fabric tensor.
For more general (and realistic) contact models,
the set $\{\Alphafab\}$ must also include the
volumetric strain and the preconsolidation stress $p_{\text{c}}$.
\par
As a further finding, the
functions $\mathbf{F}_{1}$ and $F_{2}$ in
Eqs.~(\ref{eq:sigmagen})--(\ref{eq:psigen})
are primarily associated with
the frictional forces and sliding among particles,
and these functions are assumed to depend
on the force-related variables $\{\Alphastruct\}$,
which are taken as a single tensor $\Alphastruct$.
\par
In adopting the results of Section~\ref{sec:dem},
the general
Eqs.~(\ref{eq:psiderivs})--(\ref{eq:psigen})
are assumed to have the following dependencies:
\begin{equation}\label{eq:variables}
  \begin{aligned}
  \Alphaset
  &\rightarrow \left\{ \{ \Alphafab \},\Alphastruct\right\}
  \\
  \Erev(\Varepsilon,\Alphaset)
  &\rightarrow
  \Erev\big(\Varepsilon,\{\Alphafab\}\big)
  \\
  \mathbf{F}_{1}(\Alphaset)
  &\rightarrow\mathbf{F}_{1}(\Alphastruct)
  \\
  F_{2}(\Alphaset)
  &\rightarrow F_{2}(\Alphastruct)
  \\
  \left\{\boldsymbol{\chi}\big(\Varepsilon,\Alphaset;\Xibold\big)\right\}
  &\rightarrow
  \left\{
  \Chifab\big(\varepsilon,\{\Alphafab\}\big),\,
  \Chistruct(\varepsilon,\Alphastruct ; \Xibold)
  \right\}
  \\
  \Xibold
  &\rightarrow
  \Alphadotstruct \big/ |\Alphadotstruct|
  \end{aligned}
\end{equation}
With the small set of fabric
variables $\{\Alphafab\}$ that is proposed
in Section~\ref{sec:dem}, the stiffness modulus
$\Erev$ is reduced to a function of strain
$\Varepsilon$ and variables $\{\Alphafab\}$.
The modulus $\Erev$
is also shown to behave as a conventional gradient of stress.
On the other hand, the simulations show that
$\Chistruct$ must be treated as a generalized derivative
that depends on direction $\Xibold$
of $\Alphadotstruct$.
The norm $|\cdot|$ in the definition of $\Xibold$ is taken
as the Frobenius norm.
\par
With these dependencies,
the dissipation rate in Eq.~(\ref{eq:balance}) results
from changes in both the fabric and structure variables:
\begin{equation}\label{eq:dissipform}
    \Chiset\cdot\Alphadotset
    =
    \Chifab\cdot\{\Alphadotfab\}
    \;+\;
    \Chistruct\cdot\Alphadotstruct
\end{equation}
where the stress-like fabric and structure variables,
$\Chifab$ and $\Chistruct$, are the derivatives of the free energy
in Eq.~(\ref{eq:psigen}):
\begin{equation}
  \label{eq:chis}
  \begin{aligned}
    \Chifab\big(\varepsilon,\{\Alphafab\}\big)
    &\:=\:
    -
  \int_{\Varepsilono}^{\Varepsilon}\!
  d\Varepsilon^{\prime\prime}
  \cdot\!  \int_{\Varepsilono}^{\Varepsilonpp}
    \frac{\partial\,\Erev
    \big(\Varepsilonp,\{\Alphafab\}\big)}{\partial\{\Alphafab\}}
  \cdot
  d\Varepsilon^{\prime}
  \\
  \Chistruct(\varepsilon,\Alphastruct;\Xibold)
    &\:=\:
  -
  \int_{\Varepsilono}^{\Varepsilon}\!
  \frac{\partial\mathbf{F}_{1,\Xibold}(\Alphastruct)}
       {\partial\Alphastruct}
  \cdot d\Varepsilon^{\prime}
  \:-\:
  \frac{\partial F_{2,\Xibold}(\Alphastruct)}
       {\partial\Alphastruct}
  \end{aligned}
\end{equation}
in which we have applied the Leibniz rule to place
differentiations inside their integrals.
The reversible modulus
$\Erev$ is shown to depend smoothly on fabric $\{\Alphafab\}$
(Section~\ref{sec:dem}), so that
the derivative in Eq.~(\ref{eq:chis}\textsubscript{1})
is the conventional gradient
$\Nabla_{\{\Alphafab\}}\Erev$.
By quantifying this gradient, we are able to
determine the contribution of modulus coupling~---
a non-zero derivative
$\partial\Erev/\partial\{\Alphafab\}$ in
Eq.~(\ref{eq:chis}\textsubscript{1})~---
to the total dissipation.
We accept the possibility, however,
that $\mathbf{F}_{1}$ and $F_{2}$
are not differentiable,
so that G\^{a}teaux derivatives
$\Nabla_{\Alphastruct}[\Xibold]\mathbf{F}_{1}$
and $\Nabla_{\Alphastruct}[\Xibold]F_{2}$
must be applied in Eq.~(\ref{eq:chis}\textsubscript{2}).
\par
The principles presented thus far are primarily concerned with
the internal energy $\psi$ and a bookkeeping of
its constituent parts.
In the remainder of the paper,
attention is shifted to the four variables~---
$\{\Alphafab\}$, $\Alphastruct$,
$\Chifab$ and $\Chistruct$~---
none of which can be directly controlled,
either in
laboratory experiments or with DEM simulations.
Quantifying these objects nevertheless becomes possible
by supplementing the above principles
with additional information,
either as a conjectured macro-scale
form of the energy function $\psi$
or as a conjectured micro-scale origin
of $\Alphastruct$ and $\Chistruct$.
Two proposals are now considered,
each advanced from a different viewpoint:
a macro-scale perspective, and one based on micro-mechanics.
Both proposals share the same fabric variables $\{\Alphafab\}$
but differ in the structure tensor $\Alphastruct$.
Because of this difference,
we distinguish ``Str'' and ``str''
variables, $\AlphaStruct$ and $\Alphastruct$,
for the macro- and micro-scale approaches,
along with their macro and micro complements
$\ChiStruct$ and $\Chistruct$.
\subsection{Macro-scale approach}\label{sec:macro1}
In this approach,
information is supplied as an assumed form
of the Helmholtz energy $\psi$,
thereby allowing
one to quantify the internal variable $\AlphaStruct$
and its dissipative complement $\ChiStruct$ along a load path
(here, we use the capitalized ``Str'').
We adopt the proposal of
Collins and colleagues \cite{Collins:1997a,Collins:2002a,Collins:2005b}
and view $\AlphaStruct$
and $\Varepsilon-\AlphaStruct$ as the irreversible 
and reversible strains.
Energy $\psi$ is assumed the sum of two parts,
\begin{equation}\label{eq:psimacro1}
    \psi\big(\Varepsilon,\{\Alphafab\},\AlphaStruct\big)
    \;=\;
    \psi_{1}\big(\{\Alphafab\},\Varepsilon-\AlphaStruct\big)
    \;+\;
    \psi_{2}\big(\AlphaStruct \big)
\end{equation}
where $\psi_{1}$ depends on the reversible strain
$\Varepsilon-\AlphaStruct$ and on the fabric $\{\Alphafab\}$,
\begin{equation}\label{eq:psi1a}
  \psi_{1}
  =
  \int_{\Varepsilono-\,\AlphaStructo}%
       ^{\Varepsilon-\,\AlphaStruct}\!\!\!
  d(\Varepsilon-\AlphaStruct)^{\prime\prime}
  \cdot\!
  \int_{\Varepsilono-\,\AlphaStructo}%
      ^{\Varepsilonpp\!-\,\AlphaStructpp}\!\!\!\!\!
  \Erev \big(\Varepsilon,\{\Alphafabp\}\big)
  \cdot
  d(\Varepsilon-\AlphaStruct)^{\prime}
  \;+\;
  \Sigmao
  \cdot
  (\Varepsilon-\AlphaStruct)
\end{equation}
so that $\psi_{1}$ is a double integration of the reversible
strain along the loading path.
The stress is derived from Eq.~(\ref{eq:psiderivs}\textsubscript{1}),
which is also as a function of the reversible strain
$\Varepsilon-\AlphaStruct$:
\begin{equation}\label{eq:strain1}
  \Sigmabold
  (\Varepsilon,\{\Alphafab\},\AlphaStruct)
  =
  \int_{\Varepsilono}^{\Varepsilon}
  \Erev \big(\Varepsilon,\{\Alphafabp\}\big)
  \cdot
  d(\Varepsilon-\AlphaStruct)^{\prime}
  \;+\;
  \Sigmao
\end{equation}
With both equations,
the integrand $\Erev (\{\Alphafab\})$
can be determined from DEM simulations using the methods described in
Section~\ref{sec:moduli} and \ref{sec:measuredem}.
\par
As a first step in quantifying $\AlphaStruct$,
expand the double-integral in Eq.~(\ref{eq:psi1a})
as four component integrals,
\begin{equation}\label{eq:psi1b}
  \begin{aligned}
     \psi_{1}
     =&
     \int_{\Varepsilono}^{\Varepsilon}\!\!\!
  d\Varepsilonpp
  \cdot\!\!
  \int_{\Varepsilono}^{\Varepsilonpp}\!\!\!\!
  \Erev \big(\Varepsilon,\{\Alphafabp\}\big)
  \cdot
  d\Varepsilon^{\prime}
  -
  \int_{\Varepsilono}^{\Varepsilon}\!\!\!
  d\Varepsilonpp
  \cdot\!\!
  \int_{\Alphastructo}^{\AlphaStructpp}\!\!\!\!\!
  \Erev \big(\Varepsilon,\{\Alphafabp\}\big)
  \cdot
  d\AlphaStructp
  \\
  &-
  \int_{\AlphaStructo}^{\AlphaStruct}\!\!\!\!
  d\AlphaStructpp
  \cdot\!\!
  \int_{\Varepsilono}^{\Varepsilonpp}\!\!\!\!\!
  \Erev \big(\Varepsilon,\{\Alphafabp\}\big)
  \cdot\!
  d\Varepsilon^{\prime}
  +
  \int_{\Alphastructo}^{\AlphaStruct}\!\!\!\!
  d\AlphaStructpp
  \cdot\!\!
  \int_{\AlphaStructo}^{\AlphaStructpp}\!\!\!\!\!\!
  \Erev \big(\Varepsilon,\{\Alphafabp\}\big)
  \cdot
  d\AlphaStructp
  \\
  &+\;
  \Sigmao
  \cdot
  (\Varepsilon-\AlphaStruct)
  \end{aligned}
\end{equation}
noting that
the integrals on the first line are functions
of $\Varepsilon$; whereas
integrals on the second line are functions of $\AlphaStruct$.
Comparing Eqs.~(\ref{eq:psigen}), (\ref{eq:psimacro1}), and
(\ref{eq:psi1b}),
the second integral in Eq.~(\ref{eq:psigen})
coincides with the second integral in Eq.~(\ref{eq:psi1b}):
\begin{equation}
  \mathbf{F}_{1}\big(\AlphaStruct\big)
  =
  -
  \int_{\AlphaStructo}^{\AlphaStruct}\!\!\!
  \Erev \big(\Varepsilon,\{\Alphafab\}\big)
  \cdot
  d\AlphaStructp
\end{equation}
so that at points along the load path,
\begin{equation}
  \dot{\mathbf{F}}_{1}\big(\AlphaStruct\big)
  =
  -\Erev \big(\Varepsilon,\{\Alphafab\}\big)
  \cdot
  \AlphadotStruct
\end{equation}
By using DEM results to find $\mathbf{F}_{1}$ along
the path, one solves rate $\AlphadotStruct$
and the cumulative
$\AlphaStruct$:
\begin{equation}\label{eq:alphastruct1}
\begin{aligned}
  \AlphadotStruct
  &=
  -\left(\Erev \big(\Varepsilon,\{\Alphafab\}\big)\right)^{-1}\!\!
  \cdot\,
  \dot{\mathbf{F}}_{1}\big(\AlphaStruct\big)
  \\
  \AlphaStruct
  &=
  -\!\int\!
  \left(\Erev \big(\Varepsilon,\{\Alphafab\}\big)\right)^{-1}\!\!
  \cdot\,
  d\mathbf{F}_{1}
  \;+\;
  \AlphaStructo
\end{aligned}
\end{equation} 
where $\Alphastructo$ is the initial value.
\par
With $\AlphaStruct$ solved along the path,
the second function $\psi_{2}$ in 
Eq.~(\ref{eq:psimacro1}) is quantified along the path by
noting that
the first integral on the right of
Eq.~(\ref{eq:psi1b}) is identical to the first integral
in Eq.~(\ref{eq:psigen}), and by
equating the corresponding parts of
Eqs.~(\ref{eq:psigen}) and~(\ref{eq:psimacro1})--(\ref{eq:psi1b})
and substituting Eq.~(\ref{eq:strain1}),
\begin{equation}\label{eq:psi2macro}
    \psi_{2}\big(\AlphaStruct \big)
    =\;
    \int_{\AlphaStructo}^{\AlphaStruct}\!\!\!
    \Sigmabold\cdot d\AlphaStruct
    \;+\;
    F_{2}\big(\AlphaStruct\big)
    \;-\;
    \Sigmao\cdot\Varepsilono
    \;+\;
    \psi_{\text{o}}
\end{equation}
in which function $F_{2}$ is also quantified from DEM results.
As discussed in Section~\ref{sec:microchi},
function $\psi_{2}$ is associated with the back-stress when applying the
constitutive Form~3 of Eq.~(\ref{eq:matrix3}).
\par
The assumed form of $\psi$ is given further meaning by considering
the stress rate $\Sigmadot$.
Applying Eq.~(\ref{eq:sigmadot}\textsubscript{1}) to
the form~(\ref{eq:psimacro1})--(\ref{eq:psi1a}),
\begin{equation}\label{eq:macrodstress}
    \Sigmadot
    =\;
    \Erev\big(\Varepsilon,\{\Alphafab\}\big)\cdot
    \left(\dot{\Varepsilon} - \AlphadotStruct\right)
    \;+\;
    \left(
    \int_{\Varepsilono-\,\AlphaStructo}%
      ^{\Varepsilon-\,\AlphaStruct}\!
      \frac{\partial\Erev\big(\Varepsilon,\{\Alphafabp\}\big)}
           {\partial\{\Alphafabp\}}
    \cdot
    d(\Varepsilon-\AlphaStruct)^{\prime}
    \right)
    \cdot
    \{\Alphadotfab\}
\end{equation}
The first term on the right is the difference
between the reversible change in stress
(the part of the stress rate
that is reversed by a reversal of the strain rate
$\Varepsilondot$)
and the irreversible change produced by rate
$\AlphadotStruct$;
the second term is the irreversible change in stress
due to rates of the internal variables $\{\Alphadotfab\}$,
a change that is
produced from the coupling
of modulus $\Erev$ and $\{\Alphafab\}$.
The principles in this section are quantified with DEM
results in Section~\ref{sec:macroresults}.
\subsection{Micro-scale approach}\label{sec:micro2}
Particle-scale methods, such as the DEM, allow the collection of
micro-scale information within a granular
RVE during its loading, including
measurements of
particle arrangements, movements, and contact forces.
Rather than adopting an \emph{a priori} form of the
Helmholtz energy (as done in the previous section),
we return to the principles of Section~\ref{sec:principles}
and the dependencies of Eq.~(\ref{eq:variables})
to gain insight into the nature of the internal
variables $\{\Alphafab\}$ and $\Alphastruct$
and their complementary forces $\{\Chifab\}$ and $\Chistruct$.
\par
Although both $\{\Alphafab\}$ and $\Alphastruct$
are associated with irreversible processes,
the former quantifies anisotropy and its relationship
to the modulus $\Erev$;
whereas the latter is a measure of the contact-level
friction between particles that produces internal dissipation.
In Sections~\ref{sec:moduli}--\ref{sec:dissipation},
we quantify the effects of fabric $\{\Alphafab\}$ on stiffness
$\Erev$ and on a small (coupling) dissipation.
This leaves the remaining task of identifying an
appropriate $\Alphastruct$ that authentically
chronicles frictional dissipation, which is more challenging.
Consequently, the author considered several candidates for
$\Alphastruct$ and its complement $\Chistruct$.
\par
From a micro-scale perspective,
these two quantities
(appearing in Eqs.~\ref{eq:dissipform} and~\ref{eq:chis})
are associated with internal force, movement, and dissipation.
The author considered the following criteria in their selection:
\begin{enumerate}
\item
$\Alphastruct$ and $\Chistruct$ must be tensors, providing
a directional character that accounts for both
volumetric and deviatoric conditions (as well as
loading conditions that are more general than conventional triaxial).
\item
$\Alphastruct$ must be a strain-like quantity
derived from frictional slips at sliding contacts,
while $\Chistruct$ must be stress-like,
derived from the corresponding frictional contact forces.
\item
To be consistent with the second law,
the dissipation product
$\Chistruct\cdot\Alphadotstruct$ must be consistently positive,
regardless of loading direction.
\end{enumerate}
\par
As detailed in \ref{sec:measuredem},
the actual dissipation rate $\Chiset\cdot\Alphadotset$
is the sum of contact-scale dissipation among sliding contacts,
which typically constitute a minority of all contacts:
\begin{equation}\tag{\ref{eq:DEMdissip}}
  \Chiset\cdot\Alphadotset
  =
  \varrhocsl\:\Esl
  \big(
  f_{i}^{\,\text{t}}\deltaslip_{i}
  \big)
\end{equation}
in which $\varrhocsl$ is the density of sliding contacts per volume,
and $\Esl(\cdot)$ is an expected (mean) value
among these sliding contacts.
In the equation,
the averaged quantities are the inner products of
tangential contact forces $\mathbf{f}^{\,\text{t}}$
and the corresponding tangential sliding rates $\Deltaslip$,
both of which are available in DEM simulations.
\par
Unfortunately,
the expected value
$\Esl( f_{i}^{\,\text{t}}\deltaslip_{i} )$
cannot be exactly expressed as the product of a force-like tensor
(derived from forces $\mathbf{f}^{\,\text{t}}$)
and a strain-like tensor
(derived from slips $\Deltaslip$).
This predicament calls for
approximate, albeit imperfect,
tensor proxies $\Chistruct$ and $\Alphastruct$.
\ref{sec:micro4}
analyzes candidates for these tensors and identifies
precursor tensors $\Chipre$ and $\Alphapre$
that satisfy the three criteria given above.
The stress-like $\Chistruct$ is defined as the first of
these precursors (Eq.~\ref{eq:pres}):
\begin{equation}\label{eq:chistr1}
  \chi^{\text{str}}_{ik} = \chi^{\text{pre}}_{ik} =
  \frac{\varrhocsl}{\text{tr}(\boldsymbol{Q})}
    \,
    \Esl \big( |\Ft| \big)
    \,
    \Esl \big( t^{\text{sl}}_{i}l_{k}\big)
\end{equation}
where $\mathbf{l}$ is the branch vector between the centers
of the contact's particles;
and $\mathbf{t}^{\text{sl}}$ is the
unit tangential vector of the frictional force,
$\mathbf{t}^{\text{sl}}=\Ft/|\Ft|$.
The denominator is the scalar trace of the fabric tensor
given in Eq.~(\ref{eq:Q}).
All quantities in Eq.~(\ref{eq:chistr1}) apply to
the subset of contacts that have reached the friction limit
of sliding.
\par
The corresponding strain-like precursor rate $\Alphadotpre$
is proposed in Eq.~(\ref{eq:pres}) as the product of
three expectations:
\begin{equation}\label{eq:alphadotpre}
  \dot{\alpha}^{\text{pre}}_{ik} =
  \Esl\big( |\Deltaslip| \big)
  \,
  \Big( \Esl\big(l_{l} l_{k}\big)\Big)^{-1}
  \Esl\big( t^{\text{sl}}_{i} l_{l}\big)
\end{equation}
Due to the correlation between the magnitudes
$|\Ft|$ and $|\Deltaslip|$ that appear in
Eqs.~(\ref{eq:chistr1}) and~(\ref{eq:alphadotpre}),
the precursor product
$\Chipre\cdot\Alphadotpre$ does not equal the full dissipation
required of the product $\Chistruct\cdot\Alphadotstruct$
(see Eq.~\ref{eq:pres}).
We conjecture, however, that the rates
$\Alphadotpre$ and $\Alphadotstruct$
\emph{share the same direction},
which allows the scaling of $\Alphadotpre$ so that
the full dissipation is attained.
Under this assumption, unit tensor
$\Xibold$ is designated as the direction of both
$\Alphadotpre$ and $\Alphadotstruct$:
\begin{equation}\label{eq:eta}
  \Xibold
  =
  \Alphadotpre
  /|\Alphadotpre|
  =
  \Alphadotstruct
  /
  |\Alphadotstruct|
\end{equation}
Combining Eqs.~(\ref{eq:balance}), (\ref{eq:dissipform}),
and~(\ref{eq:eta}) yields expressions for
$|\Alphadotstruct|$, $\Alphadotstruct$, and $\Alphastruct$:
\begin{equation}\label{eq:alphadis}
  |\Alphadotstruct|
  =
  -
  \frac{
   \dot{\psi} - \Sigmabold\cdot\Varepsilondot
   - \Chifab\cdot\{\Alphadotfab\}
   }
       {\Chistruct
        \cdot
        \Xibold}
  \,,\quad\;
  \Alphadotstruct =
  |\Alphadotstruct|
  \,\Xibold
  ,\quad\;
  \Alphastruct =
  \int
  \Alphadotstruct\,dt
\end{equation}
noting that all quantities on the right side of
Eq.~(\ref{eq:alphadis}\textsubscript{1})
are available from DEM simulations,
and that the final
Eq.~(\ref{eq:alphadis}\textsubscript{3})
accumulates increments of
$\Alphastruct$ along a simulation's
load path.
\section{\normalsize DEM modeling, reversible moduli,
and dissipation}\label{sec:dem}
\begin{figure}
  \centering
%
 \includegraphics{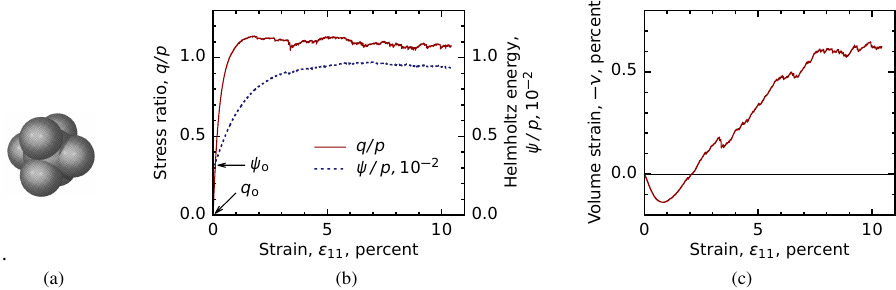}
  \caption{(a)~Seven-sphere particle.
  Results of constant-$p$ triaxial compression:
  (b)~deviatoric stress $q$ and free energy $\psi$,
  and (c)~volume change.
  \label{fig:constp}}
\end{figure}

DEM simulations were performed on
a large three-dimensional unbonded assembly
of multiple-sized particles,
with each particle modeled as a resilient,
non-breaking non-convex cluster of seven spheres
(Fig.~\ref{fig:constp}a).
Assembly characteristics are summarized in Table~\ref{table:assembly}.
Similar assemblies have been used by the author in previous works
\cite{Kuhn:2020a,Kuhn:2022a},
where the simulations closely
mimicked the behavior of clean, poorly
graded Nevada Sand.
Whereas these previous works used a Hertz-like contact model,
the present work adopts a simpler linear--frictional
contact model consisting of
linear normal and tangential springs,
with frictional sliders placed in series with the tangential springs
\cite{Suzuki:2014a}.
\par
The assembly had medium density, an isotropic fabric,
and was isotropically consolidated to a mean
stress $p$ of 100~kPa.
Because the particles are non-convex,
the average number of contacts per particle,
$2M^{\prime}/N$, was consistently greater
than the average number of contacting neighbors
per particle (coordination number, $2M/N$).
\begin{table}
  \centering\small
  \caption{\small Characteristics of the DEM assembly.%
           \label{table:assembly}}
  \begin{tabular}{lc}
    \toprule
    Characteristic & Value \\
    \midrule
    Boundaries & periodic\\
    Number of particles, $N$ & 10,648 \\
    Particle shape & sphere clusters \cite{Kuhn:2014c}\\
    Mean particle size, $\ell$ ($\approx D_{50}$) & 0.165~mm\\
    Contact stiffnesses, $k^{\text{n}}=k^{\text{t}}$ & 6000~N/m \\
    Inter-particle friction coefficient, $\mu$
      & 0.50, $\phi_{\mu}=26.5^{\circ}$\\
    Initial mean stress, $\frac{1}{3} \sigma_{kk}$ & 100~kPa\\
    Initial void ratio, $e$ & 0.731\\
    Initial particle--particle coordination number, $2M/N$ & 4.11\\
    Initial contact--contact coordination number, $2M^{\prime}/N$ & 6.71\\
    Ratio, mean (contact overlap) $/$ (particle size) & 
      $1.40\times 10^{-4}$\\
    \bottomrule
  \end{tabular}
\end{table}
\par
The simulations were
conducted with conventional triaxial conditions:
equal lateral strains
($\varepsilon_{22}=\varepsilon_{33}$) and
no shearing strain
($\varepsilon_{12}=\varepsilon_{13}=\varepsilon_{23}=0$).
The primary loading path was monotonic triaxial compression
at constant mean stress (constant-$p$).
During loading,
the periodic cell was compressed in the $x_{1}$ direction
while adjusting the lateral strains to maintain
constant mean stress.
Loading was applied in compressive strain increments of
$\Delta\varepsilon_{11}=1\times 10^{-7}$
in a nearly quasi-static manner,
such that
the average force-imbalance on the particles was about 0.0004
times the average contact force \cite{Suzuki:2014a},
and the inertia number $I$ was less than
1$\,\times\,$10\textsuperscript{$-$5} \cite{daCruz:2005a}.
A small damping is required
to maintain numerical stability in DEM simulations;
however, the energy dissipated through viscous damping was
about 0.008 times
that dissipated by contact friction.
Large inhomogeneities, such as shear bands,
were suppressed by the periodic boundaries
and by the specimen's width of about 16$\ell$.
Periodic boundaries also
abrogated the influence of the assembly's self-weight.
In any event, the self-weight of the assembly is insignificant
in comparison with the mean stress and contact stiffness:
the ratios $\gamma H/p$ and $p\ell^2/k^{\text{n}}$ were
both about 0.005 (here, $\gamma$ is the bulk unit weight,
and $H$ is the assembly height).
\par
Triaxial conditions reduce the six available components of stress
and strain to just two~--- volumetric and deviatoric~---
but even with this simplification,
the simulations demonstrate essential aspects of granular
thermomechanics.
Results
are expressed with the following pairs of generalized
stresses and strains:
\begin{equation}\label{eq:components}
\begin{alignedat}{3}
  p &= {\scriptstyle\frac{1}{3}}(\sigma_{11} + 2\sigma_{\ell})
  &\qquad
  v &= \varepsilon_{11} + 2\varepsilon_{\ell}
  \\
  q &= \sigma_{11} - \sigma_{\ell}
  &\qquad
  \epsilon &=
    {\scriptstyle\frac{2}{3}}(\varepsilon_{11} - \varepsilon_{\ell})
\end{alignedat}
\end{equation}
where $\sigma_{\ell}$ is the lateral
(confining) stress,
$\sigma_{\ell}=\sigma_{22}=\sigma_{33}$,
and $\varepsilon_{\ell}$ is the lateral strain,
$\varepsilon_{\ell}=\varepsilon_{22}=\varepsilon_{33}$.
Note that stress and strain quantities conform with the geotechnical
convention of compression being positive.
To assure that the strain rates $\dot{v}$ and $\dot{\epsilon}$
are complementary with the Cauchy stresses $p$ and $q$,
with
$\Sigmabold\cdot\dot{\boldsymbol{\varepsilon}}=
p\dot{v} + q\dot{\epsilon}$,
the rates were derived from the assembly's instantaneous velocity
gradient,
and the cumulative strains $v$ and $\epsilon$ were integrated
from their rates
as logarithmic (Hencky) strains.
\par
Figure~\ref{fig:constp} shows
stress, energy, and volume behaviors during monotonic
constant-$p$ compressive loading.
In this and subsequent figures,
we plot the engineering strain
$\varepsilon_{11}$, which serves as a time-like parameter
to track results.
The peak deviatoric stress was
$q/p=1.14$ at strain $\varepsilon_{11}=1.8\%$.
The simulation is seen to end near the critical state,
at which the deviator stress $q$ was stationary
($q/p\approx 1.07$) and changes in energy $\psi$
and volume had nearly ceased.
Although the assembly exhibited dilation during the
constant-$p$ loading,
its initial density was fairly loose:
the assembly exhibited only a small volume increase of 0.8\%
at the end of loading, and the final void ratio $e=0.742$ was
only slightly larger than the initial 0.731
These results indicate that
the initial assembly was close to the critical state $e$--$p$ line.
\subsection{%
Reversible moduli H$^{\varepsilon\varepsilon}$~--- DEM results}%
\label{sec:moduli}
In this section,
we present the results of DEM strain probes
that were used to measure the reversible modulus $\Erev$.
Section~\ref{sec:moduli2} establishes the most relevant fabric
variables $\{\Alphafab\}$ that influence these moduli.
With these variables,
we then apply homogenization methods in Section~\ref{sec:moduli3}
to arrive at estimates of $\Erev(\{\Alphafab\})$,
as these will be essential in determining
the structure variables $\Alphastruct$ and $\Chistruct$
in later sections.
\par
Past DEM studies provide ample evidence that the reversible modulus
$\Erev$ is a true gradient $\Nabla_{\Varepsilon}\Sigmabold$,
rather than merely a generalized derivative
\cite{Calvetti:2003a,Kuhn:2018c,Farahnak:2024a,Liao:2025a}.
The components of $\Erev$
were measured at 48 strains along the DEM load path of monotonic
constant-$p$ triaxial compression, using methods
described in \ref{sec:measuredem}.
At each of these strains, the system's status was fully archived
(i.e. particle positions, contact conditions, etc.),
and starting with the same archived condition,
36 strain probes were conducted within
the Rendulic $v$--$\epsilon$ plane,
with each probe having a strain magnitude
$|\Delta\Varepsilon| = 2\times 10^{-6}$
(see \cite{Calvetti:2003a,Kuhn:2018c}).
During the probes,
frictional slipping at the contacts~---
dissipation produced by changes in the internal variables
$\Alphaset$~---
was disallowed.
The four reversible moduli of the $2\times 2$ matrix 
$[\Erev ]$
were computed as a best-fit to the $p$--$q$ stress increments
for each set of 36 probes:
\begin{equation}\label{eq:defErev}
  \begin{bmatrix}
     \dot{p}^{\text{rev}}\\
     \dot{q}^{\text{rev}}
  \end{bmatrix}
  =
  \begin{bmatrix}
     \Erevn_{pv} & \Erevn_{p\epsilon}\\
     \Erevn_{qv} & \Erevn_{q\epsilon}
  \end{bmatrix}
  \begin{bmatrix}
     \dot{v}\\
     \dot{\epsilon}
  \end{bmatrix}
\end{equation}
where $\dot{p}^{\text{rev}}$ and $\dot{q}^{\text{rev}}$
are the stress increments (expressed as rates) produced
by the probe strains $\dot{v}$ and $\dot{\epsilon}$.
\par
\begin{figure}
  \centering
  \includegraphics{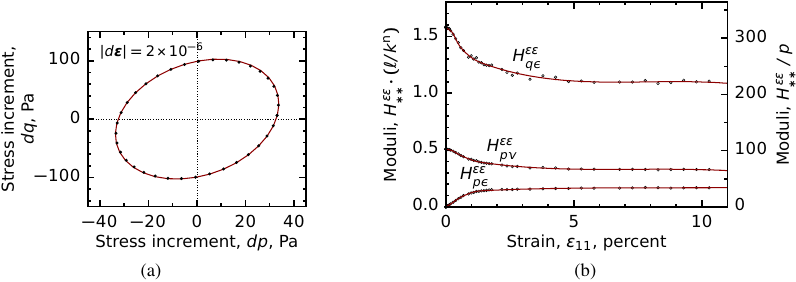}
  \caption{Moduli $\Erev$ during constant-$p$ loading,
  determined from strain probes with contact sliding suppressed:
  (a)~comparison between stress increments $d\Sigmabold$
  and those computed with the product $\Erev\cdot d\Varepsilon$; and
  (b)~components of reversible modulus 
  $\Erev$, presented in alternative dimensionless forms.
  \label{fig:moduli}}
\end{figure}

Fig.~\ref{fig:moduli}a compares 36 reversible stresses
and those computed
with Eq.~(\ref{eq:defErev}), taken
at the strain
$\varepsilon_{11}=0.6\%$.
The close fit demonstrates that modulus
$\Erev$ provides the linear relationship~---
as a true gradient~---
between stress and strain rates when irreversible changes are
suppressed.
This conclusion is consistent
with other probe simulations
\cite{Calvetti:2003a,Kuhn:2018c,Farahnak:2024a,Liao:2025a}.
\par
The four matrix components in Eq.~(\ref{eq:defErev})
comprise the reversible moduli
$\Erev$.
Fig.~\ref{fig:moduli}b shows
the evolution of these moduli during constant-$p$ loading
at 48 strains.
In keeping with Eqs.~(\ref{eq:approxE1})--(\ref{eq:approxE2}),
described below,
the moduli are presented in a dimensionless
form by dividing by the quotient $k^{\text{n}}/\,\ell$,
where $k^{\text{n}}$
is the contacts' incremental normal stiffness
and $\ell$ is the mean particle size (approximately
equal to size $D_{50}$ in geotechnical  practice).
Only three of the four moduli are shown,
since the off-diagonal moduli are found to be nearly equal.
\par
Fig.~\ref{fig:moduli}b shows that the diagonal
reversible stiffnesses,
$\Erevn_{q\epsilon}$ and $\Erevn_{pv}$,
are largest at the onset of loading.
They decrease most rapidly at strains below 2\%
and approach asymptotic values at 5\% strain.
The two stiffnesses evolve in parallel throughout loading,
with $\Erevn_{pv}$ consistently equal to 0.30--0.33
times $\Erevn_{q\epsilon}$.
Their final asymptotic stiffnesses are 0.65--0.70 of their initial values.
The figure also illustrates the evolution
of the off-diagonal reversible stiffness $\Erevn_{p\epsilon}$.
This stiffness is initially zero and increases with strain,
reaching an asymptotic normalized value of approximately
0.18 at 5\% strain.
\subsection{%
Reversible moduli H$^{\varepsilon\varepsilon}$~--- relevant fabric variables}%
\label{sec:moduli2}
We now consider the form of function
$\Erev(\Varepsilon,\{\Alphafab\})$ and
its relevant factors,
namely the strain
$\Varepsilon$ and the fabric-related internal variables
$\{\Alphafab\}$.
A soil's stiffness is known to increase with the
mean stress $p$, and the bulk modulus is often
proportional to $p^{0.5}$ \cite{Hardin:1978a}.
This dependence arises from the nonlinear
(e.g., Hertzian) stiffness of particle contacts and from an
increase in the number of contacts as 
$p$ increases
\cite{Walton:1987a,Agnolin:2007c,LaRagione:2012a,Kuhn:2014c}.
The mean stress $p=\partial\psi/\partial v$ can serve
as a proxy for the strain $\Varepsilon$ in
function $\Erev(\Varepsilon,\{\Alphafab\})$.
(Alternatively,
$p$ can replace the volumetric strain $v$ as
a state variable in the Gibbs energy,
the Legendre transformation of $\psi$ \cite{Collins:1997a}.)
\par
Besides depending on $p$,
the reversible moduli also depend on the average coordination
number or, equivalently, on the volume density of contacts
\cite{Agnolin:2007a,Goddard:1990a}.
Contact density
can be both reversible and irreversible:
some contacts that are formed with an increase in $p$ are
subsequently lost when $p$ is reduced; whereas other
contacts persist as a result of friction among the particles'
contacts, leading to irreversible particle rearrangements.
The geotechnical
preconsolidation pressure $p_{\text{c}}$ captures
such irreversible changes and is known to
influence a soil's stiffness and strength.
Accordingly,
the preconsolidation pressure
can be included as an internal variable,
insofar as its effects are not accounted by other variables.
\par
Reversible stiffness $\Erev$
decreases during deviatoric loading~---
seen in Fig.~\ref{fig:moduli}b~---
a response typical of soils
and attributed to a loss of contacts when the material is distorted.
$\Erev$ also acquires stress-induced
anisotropy during loading,
owing to the preferential disengagement of contacts
in directions of extension.
Together, these trends
distinguish granular materials as being
\emph{coupled}, with the stiffness moduli intrinsically linked to
internal fabric \cite{Hueckel:1976a,Collins:1997a}. 
\par
The general form
$\Erev(p,p_{\text{c}},\varrhoc,\boldsymbol{\beta}^{\text{fab}})$
accounts for these influences,
in which $\varrhoc=M^{\prime}/V$
is the number-density of contacts
per unit volume, and $\boldsymbol{\beta}^{\text{fab}}$ 
is a dimensionless
fabric tensor that conveys the relative numbers of contacts
in the coordinate directions.
That is, the set of fabric variables
$\{\Alphafab\}$ that carries relevant information
on the particles' arrangement includes
the three variables $p_{\text{c}}$,
$\varrhoc$, and $\boldsymbol{\beta}^{\text{fab}}$.
\par
Our simulations employed
a simple linear-frictional contact model consisting of
linear springs in series with frictional sliders
\cite{Kuhn:2020b}.
With this simple model,
the contact stiffness is independent of contact force
(unlike Hertzian contacts),
thus reducing the influence of pressure $p$ on modulus
$\Erev$, other than through its effect
on contact density $\varrhoc$.
For linear-frictional contacts, conventional
triaxial conditions,
and the simplified contact stiffnesses $k^{\text{n}}=k^{\text{t}}$
of our simulations,
we show that
the set $\{\Alphafab\}$ is further reduced to two scalars:
$\varrhoc$ and $\zeta$, the latter being a measure
of anisotropy (Section~\ref{sec:moduli3}).
In summary, the set $\{\Alphafab\}$ consists of the
following internal variables,
\begin{equation}\label{eq:Erevdep}
  \{\Alphafab\} =
  \begin{cases}
    \{p,p_{\text{c}},\varrhoc,\boldsymbol{\beta}^{\text{fab}}\}
       & \text{general contacts, general loading}
    \\
    \{\varrhoc,\boldsymbol{\beta}^{\text{fab}}\}
       & \text{linear contacts, general loading}
    \\
    \{\varrhoc,\zeta\}
       & \text{linear contacts, triaxial loading},
       k^{\text{n}}=k^{\text{t}}
  \end{cases}
\end{equation}
thus stipulating the dependencies
in Eq.~(\ref{eq:variables}).
Note that the volume strain $v$ can replace its complement $p$
in the argument list
of $\Erev$.
We now present analyses that quantify the dependency
of $\Erev$ on $\varrhoc$ and $\zeta$
for the conditions of our simulations.
\subsection{%
Reversible moduli H$^{\varepsilon\varepsilon}$~--- homogenization estimates}\label{sec:moduli3}
\par
Standard homogenization methods provide estimates
of the reversible moduli $\Erev$
(e.g., \cite{Chang:1989b,Mehrabadi:1993a,Liao:1997a,Kruyt:2004a}).
A favorable comparison of such estimates
with the simulation results (reported below) supports our use of
$\varrhoc\boldsymbol{\beta}^{\text{fab}}$~---
the product of scalar $\varrhoc$ and tensor
$\boldsymbol{\beta}^{\text{fab}}$~---
as the basis for quantifying $\Erev$.
Besides the reversible absence of contact sliding,
this homogenization is based on four assumptions:
(1)~that the contacts' stiffnesses are linear,
(2)~that the contact stiffness $k^{n}$ is much larger than
the product of contact force and particle size,
(3)~that the particles' movements strictly conform to the mean
strain field
(the so-called Voigt assumption), and
(4)~that particles do not rotate.
Although these assumptions are certainly contravened
during granular deformation,
they are applied here to develop a qualitative basis for $\Erev$,
to which we later apply numeric prefactors that bring this basis
into quantitative agreement with simulation results.
Without any prefactors,
the homogenization assumptions result in
the following approximation of the reversible stiffness:
\begin{equation}
  \Erevn_{ijkl} \approx
  \varrhoc
  \big(k^{\text{n}} - k^{\text{t}}\big)
  \Ec\big(n_{i} l_{j} n_{k} l_{l} \big)
  +
  \varrhoc
  k^{\text{t}}
  \delta_{ki}
  \Ec\big( l_{j} l_{l} \big)
\end{equation}
where $\varrhoc=M^{\prime}/V$ is the contact density;
$\Ec(\cdot)$ is the expected (mean) value of a set
of contact quantities (in this case, of the contacts'
products $n_{i} l_{j} n_{k} l_{l}$);
$k^{\text{n}}$ and $k^{\text{t}}$
are the linear normal and tangential
contact stiffnesses;
$\mathbf{l}$ is a contact's branch vector, connecting
the centers of the two particles;
$\mathbf{n}$ is the contact's normal vector;
and $\boldsymbol{\delta}$ is the Dirac identity.
\par
To further simplify the analysis, we assume
that the normal and tangential contact stiffnesses are
equal, $k^{\text{n}}=k^{\text{t}}\equiv k$,
consistent with the simulations herein.
We finally assume that
the length of branch vectors between particle
centers is uncorrelated with the vectors' orientations.
The stiffness is then approximated as
\begin{equation}\label{eq:approxE1}
  \Erevn_{ijkl} \approx
  k\ell^{2}
  \varrhoc
  \delta_{ik}
  \beta^{\text{fab}}_{lj}
\end{equation}
where $\ell$ is the mean length of branch
vectors $\mathbf{l}$ (serving as a proxy for particle size),
and $\boldsymbol{\beta}^{\text{fab}}$ is a fabric tensor
with unit trace:
\begin{equation}\label{eq:defs1}
  \ell =
  \big(\Ec(
  l_{k}l_{k})
  \big)^{1/2}
  ,\quad
  \beta^{\text{fab}}_{lj} =
  \Ec\big( m_{l}m_{j} \big)
\end{equation}
in which $\mathbf{m}$ is the unit vector aligned
with a contact's branch vector,
$\mathbf{m}=\mathbf{l}/|\mathbf{l}|$.
\par
In estimate~(\ref{eq:approxE1}),
the reversible stiffness
$\Erev$
is a function of the tensor-valued product
$\varrhoc\boldsymbol{\beta}^{\text{fab}}$, so that
the dependency
$\Erev(\varrhoc\boldsymbol{\beta}^{\text{fab}})$
is a simplification of the originally form
$\Erev(\Varepsilon,\{\Alphafab\})$.
For the triaxial conditions considered in the paper,
approximation~(\ref{eq:approxE1}) is further simplified:
stiffness $\Erev$ is a $2\times 2$ matrix that depends
only on two scalar variables,
\begin{equation}\label{eq:approxE2}
  \begin{bmatrix}
     \dot{p}^{\text{rev}}\\
     \dot{q}^{\text{rev}}
  \end{bmatrix}
  =
  \left[\Erev \big(\{\Alphafab\}\big)\right]
  \begin{bmatrix}
     \dot{v}\\
     \dot{\epsilon}
  \end{bmatrix}
  \approx
  k \ell^{2} \varrhoc
  \begin{bmatrix}
     \frac{1}{9} & \frac{1}{3}\zeta\\
     \frac{1}{3}\zeta & \frac{1}{2}(1+\zeta)
  \end{bmatrix}
  \begin{bmatrix}
     \dot{v}\\
     \dot{\epsilon}
  \end{bmatrix}
\end{equation}
where $\zeta$ is the unitless measure of fabric anisotropy,
$\zeta=\beta^{\text{fab}}_{11}-\beta^{\text{fab}}_{22}=\beta^{\text{fab}}_{11}-\beta^{\text{fab}}_{33}$
(note that matrix $\boldsymbol{\beta}^{\text{fab}}$
has unit trace).
\par
In summary, the fabric set $\{\Alphafab\}$
is reduced to two scalars, $\varrhoc$ and $\zeta$,
for the particular conditions of our simulations.
Furthermore,
the structure tensor $\Alphastruct$ is reduced
to two generalized components~---
volumetric and deviatoric~---
for conventional triaxial conditions.
The full set of internal variables $\Alphaset$, therefore,
is composed of four components:
\begin{equation}
  \Alphaset = \big\{ \{\Alphafab\},\Alphastruct \big\}
            = \big\{ \{\varrhoc,\zeta\},
            \alpha^{\text{str}}_{v},\alpha^{\text{str}}_{\epsilon} \big\}
\end{equation}
where $\alpha^{\text{str}}_{v}$ and $\alpha^{\text{str}}_{\epsilon}$
are the volumetric and deviatoric components of tensor $\Alphastruct$.

\par
\begin{figure}
  \centering
  \includegraphics{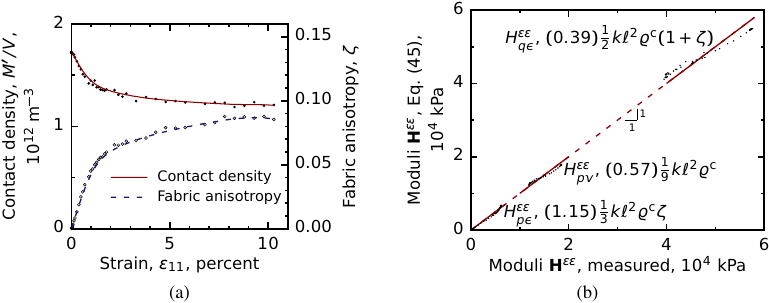}
  \caption{Fabric and reversible moduli
  at 48 strains during constant-$p$ loading:
  (a)~contact density $\varrhoc=M^{\prime}/V$ and fabric anisotropy
      $\zeta=\beta^{\text{fab}}_{11}-\beta^{\text{fab}}_{22}$,
  and
  (b)~comparison of the reversible moduli $\Erev$
  predicted with Eq.~(\ref{eq:approxE3}) and the moduli measured
  in DEM simulations.  Factors 0.39, 0.57, and 1.15 are multiplied
  by the predictions of Eq.~(\ref{eq:approxE2}) to approximate
  the simulation data.
  The red line represents perfect conformance between estimated
  and measured values.
  \label{fig:fabric1}}
\end{figure}
The evolution of contact density and fabric anisotropy
during constant-$p$ loading
is shown in Fig.~\ref{fig:fabric1}.
Contacts are lost during the first 5\% of strain
in the assembly of sphere clusters
(Fig.~\ref{fig:fabric1}a),
a result that is consistent with observations for medium dense
disk and sphere assemblies \cite{Thornton:2000a,Kruyt:2014a}.
Fabric anisotropy, quantified with parameter $\zeta$,
is initially zero (isotropic)
at the start of loading and increases until the critical state
is reached.
This behavior is similar to that observed in loose assemblies
of disks \cite{Kruyt:2016a,Zhao:2020a}
and stems from the preferential creation of contacts
in the $\varepsilon_{11}$ compression direction
and the preferential separation of contacts in the
extension directions $\varepsilon_{22}$ and $\varepsilon_{33}$
\cite{Ma:2006a,Kuhn:2010a}.
\par
Because of the homogenization assumptions, the
approximations of
Eqs.~(\ref{eq:approxE1}) and~(\ref{eq:approxE2}) are known
to overestimate the stiffness of granular
assemblies \cite{Liao:2000a,Kuhn:2003h,Kruyt:2004a}.
The moduli $\Erev$, however,
closely follow the trends predicted by these estimates.
Moreover,
by applying multiplicative prefactors to
the moduli $\Erevn_{pv}$,
$\Erevn_{p\epsilon}$, and $\Erevn_{q\epsilon}$
in Eq.~(\ref{eq:approxE2}),
the adjusted moduli achieve reasonable \emph{quantitative}
agreement with the measured values.
Figure~\ref{fig:fabric1}b shows that the factors
0.57, 0.39, and 1.15, when multiplied by
$\Erevn_{pv}$,
$\Erevn_{p\epsilon}$, and $\Erevn_{q\epsilon}$,
respectively, closely approximate
the measured values.
Discrepancies in the figure~---
differences between the measured dots and
the 1:1 line of perfect conformance~---
are likely due to $\Erev$ being affected by other aspects
of fabric not considered in the estimates.
Henceforth,
we estimate the moduli $\Erev(\{\Alphafab\})$ as
\begin{equation}\label{eq:approxE3}
  \begin{bmatrix}
     \dot{p}^{\text{rev}}\\
     \dot{q}^{\text{rev}}
  \end{bmatrix}
  =
  \left[\Erev \big(\{\Alphafab\}\big)\right]
  \begin{bmatrix}
     \dot{v}\\
     \dot{\epsilon}
  \end{bmatrix}
  \approx
  k \ell^{2} \varrhoc
  \begin{bmatrix}
     (0.57)\frac{1}{9} & (1.15)\frac{1}{3}\zeta\\
     (1.15)\frac{1}{3}\zeta & (0.39)\frac{1}{2}(1+\zeta)
  \end{bmatrix}
  \begin{bmatrix}
     \dot{v}\\
     \dot{\epsilon}
  \end{bmatrix}
\end{equation}
and we use these estimates
in the next sections to compute the energy dissipation
and stress effects
due to the coupling of stiffness
$\Erev$ and fabric variables $\varrhoc$ and $\zeta$.
\subsection{Dissipation}\label{sec:dissipation}
\begin{figure}
  \centering
  \includegraphics{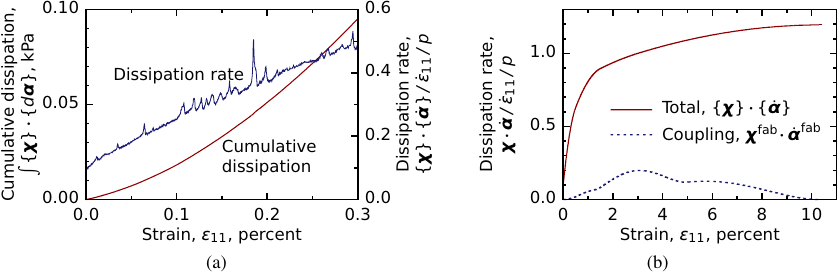}
  \caption{Frictional dissipation during constant-$p$ triaxial
           compression:
           (a)~cumulative dissipation (red) and dissipation rate (blue)
               at small strains,
           (b)~total dissipation rate (red) and the dissipation rate
               due to coupling (blue) of the reversible modulus
               $\Erev$ and the internal fabric $\Alphafab$. 
  \label{fig:dissip1}}
\end{figure}

Figure~\ref{fig:dissip1}a shows both the cumulative dissipation
$\int\!\Chiset\cdot \{d\boldsymbol{\alpha}\}$
and the dissipation rate
$\Chiset\cdot \Alphadotset$ during early strains of
constant-$p$ loading.
The dissipation rate $\Chiset\cdot \Alphadotset$
is expressed relative to the compressive strain rate
$\dot{\varepsilon}_{11}$
and we express this rate in unitless form
by dividing by the mean stress
$p=100$~kPa.
The plot ignores the prior
dissipation during the
assembly's compaction
(i.e., before the start of triaxial compression),
as the earlier dissipation is aggregated in
the initial $\psi_{\text{o}}$.
\par
At small strains, Fig.~\ref{fig:dissip1}a
shows that dissipation 
occurs \emph{at the start of loading} and that
the dissipation rate increases as loading proceeds.
The unusual early dissipation
indicates that the prior compaction process,
although conducted isotropically,
placed some of the assembly's contacts at the frictional limit,
poised to slip at the onset of loading.
As such, an elastic domain does not exist for the material
with constant-$p$ loading.
The dissipation rate in Fig.~\ref{fig:dissip1}a
increases with strain,
manifesting an increasing background of frictional sliding
among the particles.
\par
Superposed on this background dissipation,
continual spikes~--- both small and large~--
are seen in the dissipation rate,
which are caused when one or 
several contacts (among the few thousand of sliding contacts in
our simulations)
suddenly commence or cease sliding in abrupt transitions
between the sliding and non-sliding conditions.
Such fluctuations in dissipation and stress are
characteristic of granular materials
\cite{Roux:2003a,Kuhn:2019a,Baro:2021a}
and are also responsible for the raspy appearance of the
plots in Fig.~\ref{fig:constp}.
The small drops in stress in Fig.~\ref{fig:constp}b
are known to originate within small regions of the assembly
(perhaps a few hundreds of particles)
undergoing sudden alterations of their particles’ arrangement
\cite{Kuhn:2019a}.
These events produce small bursts of kinetic energy while the
local rearrangement takes place \cite{Nguyen:2016a}.
Elastic energy is, in part, transferred to kinetic energy,
but this transfer to kinetic energy is a minor destination
of the lost elastic energy.
Changes in elastic energy result primarily
from brief increases in the frictional dissipation
seen in Fig.~\ref{fig:dissip1}a \cite{Kuhn:2019a}.
The raspiness in Figs.~\ref{fig:constp} and~\ref{fig:dissip1}a
underscores the discrete nature of granular deformation
and the difficulty of drawing out continuum models that
describe macro-scale trends that occur alongside an
underlying tumult at the micro-scale.
Nonetheless, the spikes in dissipation rate (Fig.~\ref{fig:dissip1}a)
are much larger than the transitory infusions of kinetic energy
and represent brief transfers between the contacts' rates
of elastic energy and frictional dissipation.
Temporal smoothing of the two rates is thus
justified in developing the material's continuum thermomechanics.
\par
The dissipation rate for extended strains is shown by
the solid line
in Fig.~(\ref{fig:dissip1}b).
Because the spikes and dips in the rate became more
pronounced at these strains,
temporal smoothing was applied using a cubic spline fit.
The dissipation rate increases during loading and
reaches a steady plateau at the critical state,
where all stress-work, $\Sigmabold\cdot\Varepsilondot$,
is expended in frictional dissipation.
\par
In the context of thermomechanics,
dissipation is attributed to changes in
the internal variables $\Alphaset$.
By accepting the abridged dependencies of Eq.~(\ref{eq:variables}),
the dissipation rate
$\Chiset\cdot\Alphadotset$
is the sum of separate contributions by the fabric and structure
variables, which produce the rates
$\Chifab\cdot\Alphadotfab$ and $\Chistruct\cdot\Alphadotstruct$
in Eqs.~(\ref{eq:dissipform}) and~(\ref{eq:chis}).
The former derives from the coupling of
modulus $\Erev$ with fabric $\{\Alphafab\}$;
the latter originates from the rates of
functions $\mathbf{F}_{1}(\Alphastruct)$ and $F_{2}(\Alphastruct)$.
Because we have quantified the dependence of
modulus $\Erev$ on the variables variables $\{\Alphafab\}$~---
specifically, the contact density $\varrhoc$
and anisotropy $\zeta$~---
the dissipation associated with elastic–inelastic
coupling can be isolated and evaluated.
Although this coupling is expected to be
a minor contribution to dissipation,
the rate $\Chifab\cdot\Alphadotfab$ is assessed
by applying Eqs.~(\ref{eq:chis}\textsubscript{1}), in which
the equation's integrand
is quantified with Eq.~(\ref{eq:approxE3}) by computing the
derivatives of its four elements with respect to
parameters
$\varrhoc$ and $\zeta$.
\par
The dashed line in Fig.~\ref{fig:dissip1}b
shows that the dissipation associated
with stiffness-coupling is small, but it is a non-negligible
part of the system's total dissipation~---
at most, 7\% of the total dissipation rate.
Therefore, accounting for 
the coupling-related dissipation is necessary
when evaluating the
structure variable $\Alphastruct$ in both the
macro- and micro-scale approaches
of Sections~\ref{sec:macro1} and~\ref{sec:micro2},
as discussed below.
\subsection{Directional probes with friction}%
\label{sec:probes2}
The probes in Section~\ref{sec:moduli} were used for measuring
the reversible moduli $\Erev$.
In the remainder of the paper,
a separate set of probes measures irreversible effects,
and unlike the previous probes,
contacts are now allowed to slide with friction
(as in monotonic constant-$p$ loading).
At each of 48 states along
this constant-$p$ path,
the system's state was archived and, starting from each archived state,
at least 80 probes were conducted in different strain directions within
the Rendulic plane.
Each of the 48 states corresponds a starting set of the
variables $\Varepsilon$, $\{\Alphafab\}$, and $\Alphastruct$
from which probes were initiated.
These probes are used to explore
the directional nature of dissipation and to interpret
the structural variable $\Alphastruct$
from both macro- and micro-scale
viewpoints.
\subsection{Direction-dependent dissipation~--- probe results}%
\label{sec:irrev}
\begin{figure}
  \centering
  \includegraphics{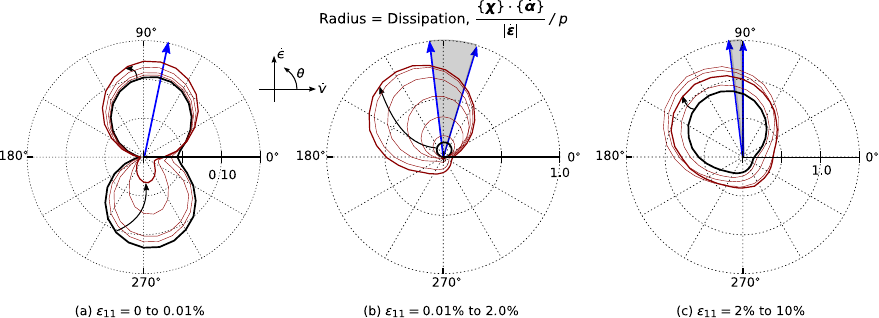}
  \caption{Polar plots of dissipation rates for strain-probes
           within the Rendulic plane:
           angles $\theta$ are strain directions
           $\tan^{-1}(\dot{\epsilon}/\dot{v})$, and
           radii are normalized dissipations.
           Each line represents 80 probes.
           Three strain ranges are shown, and
           curved arrows show trends within each range.
           Heavier black lines are results
           at the start of each range.
           Blue arrows (and shaded angles)
           show the directions of strain during prior monotonic loading.
           \label{fig:PolarDis}}
\end{figure}
%
\par
The directional nature of
dissipation is depicted in
Fig.~\ref{fig:PolarDis} for three ranges of strain.
In the figure,
scalar dissipation rates
$\Chiset\cdot\Alphadotset$
are represented by radial
distances for probes in different
$\dot{v}$-$\dot{\epsilon}$ strain directions.
Dissipation rates are shown
in unitless, normalized form
by dividing by the strain magnitude $|\Varepsilondot|$
and pressure $p$.
Curved arrows indicate trends for the advancing monotonic strain.
Blue arrows show the corresponding
directions of strain during the preceding monotonic
constant-$p$ loading.
These monotonic directions are primarily
compressive (about 90\textdegree, for
positive deviatoric strain rate $\dot{\epsilon}$),
with tilting to the left or right for dilation and contraction
(negative and positive $\dot{v}$), respectively.
\par
Results at the start of loading
(Fig.~\ref{fig:PolarDis}a) are unusual,
if one is accustomed to elastoplastic models
in which an elastic
domain is encompassed within a yield surface.
No such elastic domain exists in our DEM simulations:
dissipation occurs in all directions, even at initial loading,
regardless of whether the probe is one of triaxial compression
(upward, in the figure) or triaxial extension (downward).
This result at zero strain
indicates that the initial consolidation process~---
a process that was entirely isotropic~---
formed a specimen in which some contacts were
at the frictional limit,
a behavior that is consistent with the initial dissipation
in Fig.~\ref{fig:dissip1}a.
At the start of constant-$p$ loading ($\varepsilon_{11}=0\%$,
the solid black-line plot in Fig.~\ref{fig:PolarDis}a),
the figure shows that these frictional contact forces were evenly
distributed in all direction (isotopic);
hence the nearly mirror symmetry about the horizontal axis of the plot.
\par
As loading proceeds, the polar plots becomes more
anisotropic, and at strains greater than $\varepsilon_{11}=0.01\%$
(Figs.~\ref{fig:PolarDis}b and~c),
the dissipation rate becomes greatly reduced for negative
(unloading)
deviatoric rates $\dot{\epsilon}$ (downward in the figure).
Fig.~\ref{fig:PolarDis}b shows results in the material's
strain-hardening regime, between strain
$\varepsilon_{11}=0.01\%$ to strains slightly past the
peak load, $\varepsilon_{11}=2\%$.
Here, the behavior conforms closer to an elastoplastic
framework: dissipation is largest
within the upper half-plane of strain
loading ($\dot{\epsilon}>0$), a result that is consistent with conventional
elastoplasticity and with other 
DEM probe studies
\cite{AlonsoMarroquin:2005b,Calvetti:2003a,Sibille:2007b,
Plassiard:2009a,Wan:2014a,Kuhn:2018c}.
The results show, however, that some dissipation
occurs in all directions of loading, 
even though dissipation rates are small
between strain angles of 270$^{\circ}$ and 360$^{\circ}$.
This finding means
that a purely elastic region (e.g., within a yield surface)
does not exist during strain hardening.
A similar result was reached by
Tamagnini et al. \cite{Tamagnini:2005a}
and the author \cite{Kuhn:2018c}.
\par
At larger strains,
this lack of elastic strain directions
is more pronounced during strain softening and at the
critical state (Fig.~\ref{fig:PolarDis}c).
Significant dissipation occurs in all strain directions~---
loading and unloading~---
and at these larger strains the probes
with the least dissipation
produced dissipation rates as large as 30\% of that occurring
in the direction of monotonic constant-$p$ loading.
These anomalous results are a consequence of the erratic
nature of the particles' movements,
a matter that has been explored by the author
\cite{Kuhn:2003d,Kuhn:2016b}
and is further investigated in the next section. 
\subsection{Micro-movements~--- unusual results during
            strain-reversal}\label{sec:reversals}
If one is accustomed to the conventional elastoplastic
notions of yield surfaces and plastic potentials,
the results of particle-scale simulations can be surprising.
The lack of an elastic domain in Fig.~\ref{fig:PolarDis}
is one such anomaly.
Three sets of micro-scale results will
shed light on this macro-scale behavior.
\par
\begin{figure}
  \centering
  \includegraphics{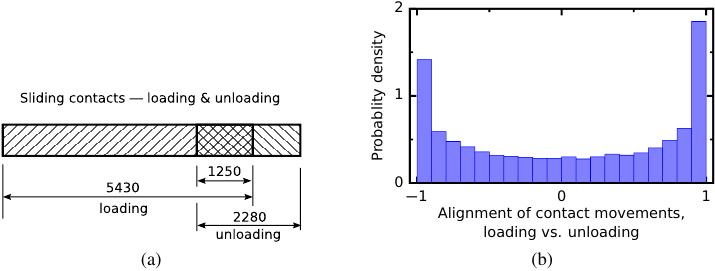}
  \caption{Micro-scale results for forward and reversed probes
           at the critical state:
           (a)~Venn diagram of the sliding contacts in the
           two probes; and
           (b)~direction cosines (alignments) of movements
           $\{\dot{\BDelta}\}$ of forward and reversed
           loadings of all 23,700 contacts.
           The direction cosines are computed as the
           inner product
           $(\dot{\BDelta}^{\text{loading}}\cdot\dot{\BDelta}^{\text{unloading}})/ |\dot{\BDelta}^{\text{loading}}||\dot{\BDelta}^{\text{unloading}}|$.
           \label{eq:anomaly}}
\end{figure}

To begin,
Fig.~\ref{eq:anomaly}a
gives results of two strain probes,
with both probes starting from the same condition,
in this case at the critical state,
but with the probes proceeding in opposite directions:
a forward probe in the same direction
as the preceding monotonic loading
($\dot{\epsilon}>0$, $\dot{v}=0$),
and a reversed, unloading probe in the opposite direction
($\dot{\epsilon}<0$, $\dot{v}=0$).
The two probes appeared as directions 90\textdegree\ and
270\textdegree\ in Fig.~\ref{fig:PolarDis}.
The results in Fig.~\ref{eq:anomaly}a
demonstrate that a reversal of the macro-scale strain
does not reverse the micro-scale movements.
The Venn diagram in Fig.~\ref{eq:anomaly}a shows that
during the forward probe, 5430
contacts were sliding
(about 23\% of all $M^{\prime}=23700$ contacts).
As expected, fewer contacts slide during unloading
(only 2280 contacts), but this 9.6\% of all 23700 contacts
represents a substantial fraction~--- a fraction responsible
for the dissipation seen in the lower half of the
polar plot in Fig.~\ref{fig:PolarDis}c.
Also unusual is that many of the same contacts that
were sliding during forward loading \emph{continue to slide}
during reverse unloading (1250 contacts,
or 23\% of the original 5430 sliding contacts).
\par
Figure~\ref{eq:anomaly}b
provides further illustration that contact movements are not reversed
by a reversal of the strain direction.
Here, we consider the 23,700 contacts at the critical
state and compare the two sets of contact movements
$\{\dot{\BDelta}\}$ in probes of forward (loading)
and reversed (unloading) directions.
The figure is a histogram of the
alignments (directional cosines) of individual movements.
Values of $-1$ apply to contacts whose movement directions
are fully reversed;
values of $1$ are contacts that continue to move in their
same (loading) directions.
Although one might expect a preponderance of $-1$ values,
Fig.~\ref{eq:anomaly}b shows that there is little correlation between
the contact movements during loading and unloading,
and many contacts \emph{continue to move in their loading
directions} (i.e., with values close to 1).
This result is obtained even though a strain reversal
is observed at the assembly's boundaries.
\par
The results of Figs.~\ref{eq:anomaly}a
and~\ref{eq:anomaly}b are taken at the critical state
(at strain $\varepsilon_{11}=10\%$), where one would expect
the most erratic, disordered movements \cite{Kuhn:2016c}.
The same general lack of loading/unloading coherence,
however, applies at all strains,
and results at the peak load are even more pronounced.
\par
\begin{figure}
  \centering
  \includegraphics{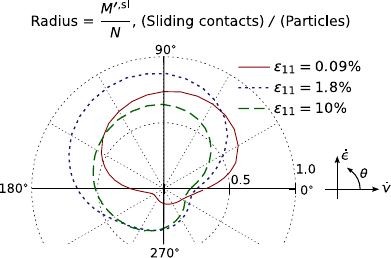}
  \caption{Polar plot for strain probes in directions
           $\theta$, showing the numbers of sliding contacts
           $M^{\prime,\text{,sl}}$. 
           Values of $M^{\prime,\text{,sl}}$ are normalized by
           dividing by the total number of particles $N$.
           \label{fig:polarslide}}
\end{figure}

Fig.~\ref{fig:polarslide} is a final illustration of the
unusual conduct among the assembly's contacts.
Each of the three polar plots correspond to a
series of strain probes that abruptly followed a
strain-state during monotonic constant-$p$ loading.
The three states are those of low strain $\varepsilon_{11}$,
the peak stress, and the critical state.
The plots are of the numbers of
\emph{sliding} contacts $M^{\prime\text{,sl}}$
for different strain-probe directions $\theta$,
divided by the number of contacts $N$.
As a reference direction,
constant-$p$ loading produced strain
directions of between 73$^{\circ}$ to 98$^{\circ}$.
The results show that
contacts slide at all strains and in all probe directions,
even though fewer contacts slide
at small strains and fewer contacts slide
for the unloading probes.
We conclude that
sliding is a certainty in granular deformation~---
occurring during loading, unloading, and in all other directions~---
producing a pervasive dissipation.
\section{\normalsize Results in a macro-scale setting}\label{sec:macroresults}
In the macro-scale setting,
variable $\AlphaStruct$ is assumed to be the irreversible strain,
and the stress rate $\Sigmadot$ is the material's response
to the reversible strain rate, equal to the difference
$\Varepsilondot-\AlphadotStruct$
(for the macro setting we use the uppercase superscript
``Str'', as $\AlphaStruct$).
The irreversible rate is also augmented by a small contribution
from the coupling
of modulus $\Erev$ and the fabric variables $\{\Alphafab\}$
(see Section~\ref{sec:macro1} and Eq.~\ref{eq:macrodstress}).
With these assumptions, DEM data is used below to
quantify the irreversible strain $\AlphaStruct$
along a constant-$p$ load path, the rate
$\AlphadotStruct$ for different probe directions along
the path,
and the evolution of the additional energy function $\psi_{2}$.
These results comprise the essential
elements of the material's constitutive framework.
We also quantify the stress-like $\ChiStruct$ and its effect
on the function $\psi_{2}$,
the latter being associated with the macro-scale back-stress.
\subsection{Macro-scale approach~---
            evolution of the irreversible strain
            $\AlphaStruct$}
\begin{figure}
  \centering
  \includegraphics{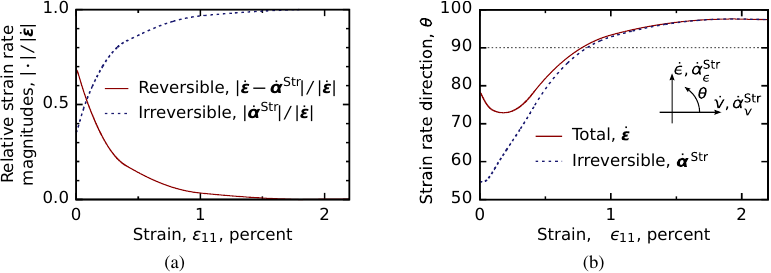}
  \caption{For the macro-scale model, strain rates during
           constant-$p$ loading:
           (a)~magnitudes of reversible and
               irreversible rates, relative to the total
               strain rate,
           (b)~directions of the total and irreversible
               strain rates.
           Angles $\theta$ greater than 90\textdegree\ are dilative;
           angles less that 90\textdegree\ are contractive.
           The peak stress occurred at 1.8\% strain.
  \label{fig:macrostrains}}
\end{figure}

The irreversible strain $\AlphaStruct$ is computed
along the constant-$p$ loading path
by applying Eq.~(\ref{eq:alphastruct1}).
Fig.~\ref{fig:macrostrains} shows the 
rates of irreversible and reversible strains for
the range of strains, 0\% to 2.2\%,
which includes the initial strain hardening and the peak stress.
The magnitudes of rates,
$|\Varepsilondot-\AlphadotStruct|$
and $|\AlphadotStruct|$ are expressed as fractions
of the total strain rate $|\Varepsilondot|$
(Fig.~\ref{fig:macrostrains}a).
For strains less than 2\%,
the full strain rate $\Varepsilondot$ is composed
of both reversible and irreversible parts,
and irreversible strain is present even at the start of loading.
As expected,
the fraction of the reversible rate decreases with strain,
and at the peak stress, nearly all further strain is irreversible,
with the magnitude of the reversible rate
being less than 0.2\% of the full strain rate.
\par
The directions of the total and irreversible strains
rates are given in Fig.~\ref{fig:macrostrains}b, shown
as angles $\theta$ relative to the volumetric
$\dot{v}$ axis of the Rendulic plane.
To put $\theta$ in perspective,
angles between $0^{\circ}$ and $180^{\circ}$
correspond to triaxial compression ($\dot{\epsilon}>0^{\circ}$);
dilation occurs between 90\textdegree\ and
270\textdegree\ ($\dot{v}<0$); and
contraction corresponds to $-90^{\circ}<\theta<90^{\circ}$.
Although its direction is not shown,
the reversible rate,
$\Varepsilondot-\AlphadotStruct$, is consistently
dilative, with a ratio $\dot{v}/\dot{\epsilon}<0$
Indeed, for constant-$p$ loading, with $\dot{p}=0$.
the sign of $\dot{v}/\dot{\epsilon}<0$ is the same
as the sign of the quotient
$-\Erevn_{p\epsilon}/\Erevn_{pv}$
in Eq.~(\ref{eq:defErev}), a value that is consistently
dilative.
\par
At the start of loading, the irreversible
strain $\AlphadotStruct$ is contractive,
which counteracts
the dilative reversible rate $\Varepsilondot-\AlphadotStruct$,
producing a total strain that is contractive
(see Fig.~\ref{fig:constp}c).
As loading continues,
the irreversible rate becomes less contractive and
transitions to dilative ($\theta>90^{\circ}$),
and in conjunction with the dilative reversible rate, causes
the material to dilate until the critical state is attained.
\subsection{Special plotting method}%
\label{sec:plotting}
We take a brief detour from DEM results to introduce the
special plotting method that is used in the remainder of the paper
to depict the results of directional probes.
Although the plotting is restricted to two-dimensional
triaxial  conditions,
it clearly exposes essential aspects of the probe results.
For example, the plots reveal whether the material's response
is sufficiently smooth
(i.e., differentiable) so that the derivative of
energy $\psi(\Varepsilon,\Alphaset)$ with respect to variable
$\Alphastruct$ is a gradient,
or, conversely, whether the response derives
from a generalized derivative, as in Eq.~(\ref{eq:Gateaux}).
Moreover, in the macro-scale setting,
the plots candidly reveal whether simulation results
are consistent with the precepts of conventional elastoplasticity.
\par
\begin{figure}
  \includegraphics{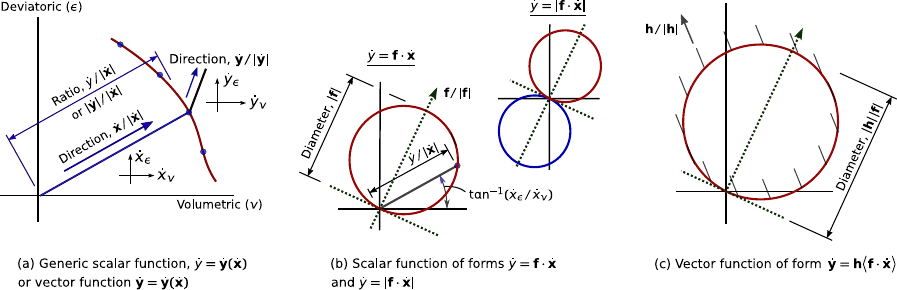}
  \caption{Plotting method for exposing the smoothness
           of a function $\dot{y}(\dot{\mathbf{x}})$
           or $\dot{\mathbf{y}}(\dot{\mathbf{x}})$: 
           (a)~general plotting method;
           (b)~special case of a smooth (i.e., \emph{differentiable})
           scalar function $y(\mathbf{x})$, in which
           $\dot{y}=\mathbf{f}\cdot\dot{\mathbf{x}}=f_{i}\dot{x}_{i}$,
           with
           $\mathbf{f}=\Nabla_{\mathbf{x}}y=
           \partial y/\partial \mathbf{x}$.
           also where
           $\dot{y}=|\mathbf{f}\cdot\dot{\mathbf{x}}|$; and
           (c)~special case of a smooth (i.e., \emph{differentiable})
           vector function $\mathbf{y}(\mathbf{x})$, in which
           the rate of function $\mathbf{y}(\mathbf{x})$
           in the direction of $\dot{\mathbf{x}}$ is in the form
           $\dot{\mathbf{y}}=\mathbf{h}\langle\mathbf{f}
           \cdot\dot{\mathbf{x}}\rangle
           =h_{i}\langle f_{j}\dot{x}_{j}\rangle$,
           with Macaulay brackets $\langle\;\rangle$.
           In~(a) and~(c), the small spars are the directions
           of vector $\dot{\mathbf{y}}$.
           \label{fig:Polarspars2}}
\end{figure}

In the two-dimensional Rendulic setting of triaxial loading,
the planar plot represents a rate-response~--- either $\dot{y}$
or $\dot{\mathbf{y}}$~--- to a vector-valued input
rate $\dot{\mathbf{x}}$: the rate
$\dot{y}(\dot{\mathbf{x}})$
of a scalar-valued function
or the rate $\dot{\mathbf{y}}(\dot{\mathbf{x}})$ of a
vector-valued function (Fig.~\ref{fig:Polarspars2}a).
Rate $\dot{\mathbf{x}}$ will represent either a strain
$\Varepsilondot$ with its two triaxial
components $\dot{v}$ and $\dot{\epsilon}$
or the strain-like rate $\Alphadotstruct$
with components $\dot{\alpha}^{\text{str}}_{v}$ and
$\dot{\alpha}^{\text{str}}_{\epsilon}$.
The direction of $\dot{\mathbf{x}}$ is the polar
angle $\tan^{-1}\dot{x}_{\epsilon}/\dot{x}_{v}$,
measured from the horizontal $v$ axis.
The radial distance from the origin is the quotient
$\dot{y}/|\dot{\mathbf{x}}|$ or
$|\dot{\mathbf{y}}|/|\dot{\mathbf{x}}|$ for scalar or
vector functions, with each quotient being
a \emph{directional derivative} in the direction of $\dot{\mathbf{x}}$
(e.g., direction $\Xibold$, when $\dot{\mathbf{x}}$ represents $\Alphadotstruct$).
For vector functions $\dot{\mathbf{y}}(\dot{\mathbf{x}})$,
the end of a radial segment is adorned
with a small spar that points in the direction of
the response $\dot{\mathbf{y}}$.
In this manner, the locus of points (and the set of spars)
allows full visualization of the function
$\dot{y}(\dot{\mathbf{x}})$ or $\dot{\mathbf{y}}(\dot{\mathbf{x}})$.
\par
In the case of a scalar function
$\dot{y}(\dot{\mathbf{x}})$
of a two-dimensional rate $\dot{\mathbf{x}}$,
if rate $\dot{y}$ is equal to an inner product
$\mathbf{f}\cdot\dot{\mathbf{x}}$,
then the locus of points is a circle
that passes through the plot's origin (Fig.~\ref{fig:Polarspars2}b).
Note that the single circle is actually two superposed
circles, with one each for positive and negative values
of $\mathbf{f}\cdot\dot{\mathbf{x}}$,
located at a 180\textdegree\ difference.
Specifically, if $\mathbf{f}$ is a true gradient of
some scalar function with respect to $\mathbf{x}$,
say $U$ with $\mathbf{f}=\Nabla_{\mathbf{x}}U$,
then the plot is a circle having the diameter
$|\mathbf{f}|$ and oriented in the direction of $\mathbf{f}$.
One instance is the stress-work
$\dot{y}=\Sigmabold\cdot\Varepsilondot$,
with the substitutions
$\dot{\mathbf{x}}\rightarrow\Varepsilondot$
and $\mathbf{f}\rightarrow\Sigmabold=\Nabla_{\Varepsilon}\psi$.
The special plot of $\dot{y}=\mathbf{f}\cdot\dot{\mathbf{x}}$
(i.e., of $\Sigmabold\cdot\Varepsilondot$)
is a circle with diameter
$|\mathbf{f}|=|\Sigmabold|$ that
is oriented in the stress direction
$\tan^{-1}f_{\varepsilon}/f_{v}=\tan^{-1}q/p$.
Indeed, if the plot of rate $\dot{\psi}$ is derived from data
in which $\Alphadotset$ is constant
but the plot is \emph{not} a circle,
then stress $\Sigmabold\neq\Nabla_{\Varepsilon}\psi$.
Readers are assured that this result was not the case and that 
Eq.~(\ref{eq:psiderivs}\textsubscript{1}) is supported by the data.
\par
A more problematic case is the dissipation rate
and its relation to
a chosen internal variable $\boldsymbol{\alpha}$.
If $\dot{y}$ represents the measured dissipation rate,
one can determine whether the rate is consistent with the
inner product of a
single dissipation stress $\Chibold$ and the
rate $\dot{\boldsymbol{\alpha}}$,
or $\dot{y}=\Chibold\cdot\dot{\boldsymbol{\alpha}}$
(see Eq.~\ref{eq:balance}).
As shown in subsequent sections,
such plots are not circles.
With these special plots,
we will conclude that $\Chibold$ is not a true gradient
$\Nabla_{\boldsymbol{\alpha}}\psi$, but instead is
a generalized derivative
$\Nabla_{\boldsymbol{\alpha}}[\Xibold]\psi$.
\par
If a \emph{vector function} $\dot{\mathbf{y}}(\dot{\mathbf{x}})$
is equal to a product $\mathbf{A}\cdot\dot{\mathbf{x}}$
($\mathbf{A}$ being a general $2\times 2$ matrix),
the plot of $|\dot{\mathbf{y}}|/|\dot{\mathbf{x}}|$
can have various shapes, but the \emph{component-wise}
plots of rows $\dot{y}_{1}/|\dot{\mathbf{x}}|$ and
$\dot{y}_{2}/|\dot{\mathbf{x}}|$ will each be a
circle
(an example is the rate $\Sigmadot$ in Eq.~\ref{eq:approxE3}).
This situation of
$\dot{\mathbf{y}}=\mathbf{A}\cdot\dot{\mathbf{x}}$
applies when $\mathbf{y}(\mathbf{x})$ is differentiable
with respect to $\mathbf{x}$, and $\mathbf{A}$ represents
the gradient of $\mathbf{y}$, or
$\mathbf{A}=\Nabla_{\mathbf{x}}\mathbf{y}$.
\par
Fig.~\ref{fig:Polarspars2}c depicts the special case
when the gradient (the matrix $\mathbf{A}$) is a dyadic product,
$\mathbf{A}\rightarrow\Nabla_{\mathbf{x}}\mathbf{y}=\mathbf{h}\otimes\mathbf{f}=h_{i}f_{j}$,
so that the response
$\dot{\mathbf{y}}=
(\mathbf{h}\otimes\mathbf{f})\cdot\dot{\mathbf{x}}
=h_{i}f_{j}\dot{x}_{j}$.
This case applies to conventional elastoplasticity,
in which $\mathbf{f}$ is in the yield direction
and $\mathbf{h}$ is the flow direction.
If so,
the special plot is a pair of circles, each passing through the origin,
with the circles oriented in the direction of $\mathbf{f}$
and having diameter $|\mathbf{h}||\mathbf{f}|$
(two separate circles, instead of superposed circles, result
from plotting the magnitude $|\dot{\mathbf{y}}|$).
The spars, representing the direction of the response 
$\dot{\mathbf{y}}$, are oriented in the direction of $\mathbf{h}$,
but are pointed in opposite directions for the two circles.
A single circle ensues when
$\Nabla_{\mathbf{x}}\mathbf{y}=\mathbf{h}\otimes\mathbf{f}$,
except that the response is zero when
$\mathbf{f}\cdot\dot{\mathbf{x}}<0$
(we use the Macaulay brackets $\langle\cdots\rangle$ to
effect this situation, with
$\dot{\mathbf{y}} 
=\mathbf{h}\langle\mathbf{f}\cdot\dot{\mathbf{x}}\rangle$).
This case is of particular interest,
since it is encountered in conventional elastoplasticity,
in which the plastic strain rate
$\dot{\varepsilon}^{\text{pl}}_{ij}
=(1/L)h_{ij}\langle C_{pqkl}f_{pq}\dot{\varepsilon}_{kl}\rangle$,
where $\mathbf{C}$ is the elastic compliance,
$\mathbf{h}$ is the flow direction,
$\mathbf{f}$ is the yield direction, and $L$ is a scalar 
measure of hardening
(see \cite{Lubliner:1990a}, noting that
$(1/L)C_{pqkl}f_{pq}$ takes the role of $\mathbf{f}$ in the
dyad $\mathbf{h}\otimes\mathbf{f}$).
\subsection{Macro-scale approach~--- conventional elastoplasticity}
We now use the probes described in Section~\ref{sec:probes2}
to investigate the relationship between the
incremental rates $\Varepsilondot$ and $\AlphadotStruct$.
In these probes,
the monotonic constant-$p$ loading is suspended and
small strain increments are imposed in different directions
$\Varepsilondot$.
Assuming that variable $\AlphaStruct$ represents
the irreversible strain,
the probes are used to evaluate the observed incremental behavior
in relation to conventional (i.e., single-mechanism) elastoplasticity
and to assess consistency of the observations
with, or their divergence from, this constitutive framework.
Specifically, we test four postulates
of conventional elastoplasticity:
\begin{enumerate}
\item
whether incremental strain directions exist with a purely
reversible response;
\item
whether a yield surface with normal direction $\mathbf{f}$
separates two tensorial half-spaces
of irreversible and reversible behaviors;
\item
whether the irreversible strain rate is
uniformly in a single flow direction $\mathbf{h}$
that is independent of the direction of strain $\Varepsilondot$; and
\item
whether the irreversible strain rate $\AlphadotStruct$
is proportional to the projected strain rate
$\mathbf{f}\cdot\Varepsilondot$.
\end{enumerate}
These postulates are expressed with the following
relationship between the rates $\Varepsilondot$ and $\AlphadotStruct$
of the total and irreversible strains:
$\AlphadotStruct=
\gamma\mathbf{h}\langle\mathbf{f}\cdot\Varepsilondot\rangle
=\gamma h_{i}\langle f_{j}\dot{\varepsilon}_{j}\rangle$,
where $\mathbf{f}$ and $\mathbf{h}$ are the
yield and flow directions at a given state
$(\Varepsilon,\Alphaset)$;
$\gamma$ is a scalar;
and $\langle\cdots\rangle$ are Macaulay brackets,
$\langle z\rangle=\text{max}\{0,z\}$.
\par
\begin{figure}
  \centering
  \includegraphics{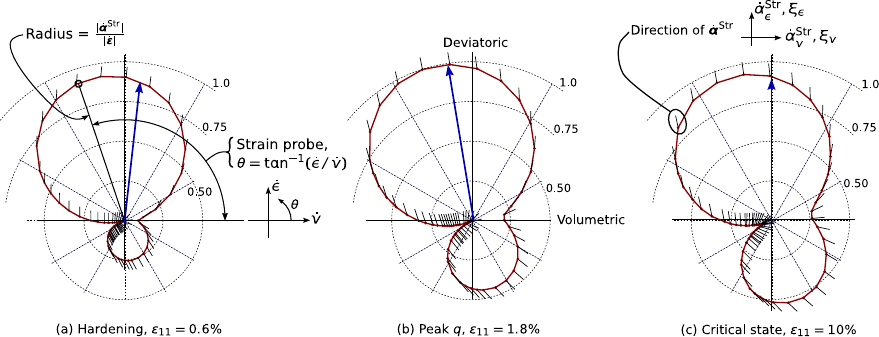}
  \caption{For the macro-scale model,
           plots of probe results,
           giving the relationship between the vector rates
           $\Varepsilondot$ and $\AlphadotStruct$
           (see Fig.~\ref{fig:Polarspars2}c for plotting method).
           Three states of the prior constant-$p$ monotonic
           loading are shown: (a)~hardening domain,
           (b)~peak deviator stress, and
           (c)~critical state.
           Angles $\theta$ correspond to strain-probe directions
           $\Varepsilondot$,
           and radial distances to the red lines give the rate ratio
           $|\AlphadotStruct|/|\Varepsilondot|$ for these
           strain directions.
           The spars point in the direction of
           rates $\AlphadotStruct$.
           The blue arrows are the direction of $\Varepsilondot$
           during the prior constant-$p$ loading.
           See Section~\ref{sec:plotting}
           and Fig.~\ref{fig:Polarspars2}c for 
           context the plotting method.
           \label{fig:macroalphaspars}}
\end{figure}

Figure~\ref{fig:macroalphaspars}
shows the results of three sets of probes, conducted
at different strain states:
during strain hardening,
at the peak deviator stress,
and at the critical state.
Each plot presents the relationship between the
input strain direction $\Varepsilondot$
and the resulting irreversible rate $\AlphadotStruct$.
Because conventional triaxial conditions involve only two
generalized strains~--- volumetric and deviatoric~---
the rates can be fully depicted in these planar plots.
By using the plotting method of Section~\ref{sec:plotting},
Fig.~\ref{fig:macroalphaspars} provides
a direct visual means of testing the four postulates.
Radial distances to the red line correspond to the ratio
$|\AlphadotStruct|/|\Varepsilondot|$, which represents the
extent to which the irreversible strain counteracts
(or, perhaps, augments) the total strain.
Joining the origin to points on the red
line gives the direction of strain rate $\Varepsilondot$.
The short spars point in the direction of the
rate $\AlphadotStruct$ (i.e., in direction $\Xibold$).
\par
With regard to the first postulate of elastoplasticity,
the plots of Fig.~\ref{fig:macroalphaspars}
never touch the origin, although cusps in the plots
are close to the origin at an angle of about 190\textdegree.
This result means that
the irreversible strain rate $\AlphadotStruct$ never truly
vanishes, and a purely reversible response does not exist for
the three strain states shown.
This conclusion is also reached with the previous
Fig.~\ref{fig:PolarDis}, showing
that dissipation occurs in all strain directions $\Varepsilondot$,
even at the onset of initial loading (at strain $\varepsilon_{11}=0$).
The particle-scale phenomena that lead to such pervasive
dissipation were described in Section~\ref{sec:reversals}.
\par
Regarding the second postulate of distinct tensorial zones
of reversible and irreversible behaviors,
Fig.~\ref{fig:macroalphaspars} shows that irreversibility
occurs in all directions.
The figure does reveal upper and lower circular halves separated
by a line of roughly 10\textdegree.
The upper circles are oriented at about 105\textdegree,
and the lower circle at about 285\textdegree, and
these angles corresponding to
the orientations of two, nearly opposing, yield directions.
\par
Contrary to the third postulate,
the spars in each plot of Fig.~\ref{fig:macroalphaspars}
do not point in a uniform direction,
meaning that the incremental irreversible strain has
multiple directions, depending on the strain direction.
In the upper parts of these plots, the direction is
87\textdegree--95\textdegree;
in the lower part it is 300\textdegree--315\textdegree;
and near the cusps, the direction transitions
between the upper and lower directions.
\par
The fourth postulate supposes that the magnitude of the irreversible
strain rate $\AlphadotStruct$
is proportional to an inner product
$\mathbf{f}\cdot\Varepsilondot$ but vanishes when the
product is negative
(vector $\mathbf{f}$ is the yield direction in strain-space).
For this assumption to be valid,
each plot in Fig.~\ref{fig:macroalphaspars} would
consist of a single circle passing through the origin.
This premise is not supported by the simulations: instead,
two near-circles are joined on one side with a blunt(er)
cusp and on the
other side by a smoother transition.
\par
In short, all four postulates of conventional (single-mechanism)
are transgressed in the simulations.
\par
The presence of two nearly circular shapes supports a
multi-mechanism model and suggests that
irreversibility can occur on both sides of a yield surface~---
whether $\mathbf{f}\cdot\Varepsilondot$ is
positive or negative
\cite{Manzari:1997a,Hashiguchi:2005a,Papadimitriou:2019a}.
Although complex, the data in Fig.~\ref{fig:macroalphaspars}
indicates that irreversibility involves
\emph{three mechanisms}, with each mechanism modeled as
an irreversible strain rate of the form
$\gamma^{(\text{i})}\mathbf{h}^{(\text{i})}
\langle\mathbf{f}^{(\text{i})}\cdot\Varepsilondot\rangle$,
with $\text{i}\in\{1,2,3\}$ and Macaulay brackets $\langle\cdot\rangle$.
The three mechanisms correspond to the plots'
large upper circles (function~1),
the smaller lower circles (function~2), and the transition regions
on the right between upper and lower circles (function~3).
The full irreversible strain
rate $\AlphadotStruct$ is the combination
\begin{equation}\label{ref:threeemech}
  \AlphadotStruct =
  \sum_{\text{i}=1,2,3}
  \gamma^{(\text{i})}\mathbf{h}^{(\text{i})}\langle
  \mathbf{f}^{(\text{i})}\cdot\Varepsilondot\rangle
\end{equation}
If vectors $\mathbf{f}^{(\text{i})}$ and
$\mathbf{h}^{(\text{i})}$ are taken as unit vectors,
then the scalars $\gamma^{\text{(i)}}$ are the maximum
rates of a mechanism's
irreversible rate $|\AlphadotStruct|$ relative to
the strain rate $|\Varepsilondot|$.
The nine parameters are then read directly from the plots
in Fig.~\ref{fig:macroalphaspars}, simply
as the circles' diameters (magnitudes $\gamma^{\text{i}}$),
and orientations (directions $\mathbf{f}^{(\text{i})}$),
and the spars' orientations (directions $\mathbf{h}^{(\text{i})}$).
In this manner, the
yield directions $\mathbf{f}^{(\text{i})}$ 
are unit vectors with orientations of
$\mathbf{f}^{(\text{1})}\approx105$\textdegree,
$\mathbf{f}^{(\text{2})}\approx285$\textdegree, and
$\mathbf{f}^{(\text{3})}\approx15$\textdegree,
with the latter vector pointing toward the right-side transition
of the two circles.
Likewise, the $\mathbf{h}^{(\text{i})}$ vectors are
unit flow vectors with (spar) orientations
$\mathbf{h}^{(\text{1})}\approx$ 87\textdegree--95\textdegree,
$\mathbf{h}^{(\text{2})}\approx$ 300\textdegree--315\textdegree, and
$\mathbf{h}^{(\text{3})}\approx$ 0\textdegree.
The scalar intensities $\gamma^{\text{(i)}}$
are the circles' diameters:
$\gamma^{\text{(1)}}=0.9$--1.0,
$\gamma^{\text{(2)}}=0.25$--0.55, and
$\gamma^{\text{(3)}}=0.1$--0.25, with all scalars being dimensionless.
With this data, Eq.~(\ref{ref:threeemech}) 
closely fits the simulation's results.
\subsection{Macro-scale approach~--- derivative $\ChiStruct$ and
function $\varphi_{2}$}%
\label{sec:microchi}
We have investigated essential elements of a macro-scale thermomechanics:
quantifying the relationship between
the fabric variables $\{\Alphafab\}$ and the reversible stiffness $\Erev$;
measuring function
$\mathbf{F}_{1}(\AlphaStruct)$
to determine the
macro-scale variable $\AlphaStruct$;
and determining the relationship between the
rates $\Varepsilondot$ and $\AlphadotStruct$.
In the context of constitutive Form~2 in
Section~\ref{sec:constituteforms},
the relationship of $\Varepsilondot$ and $\AlphadotStruct$
determines the second row of the matrix Eq.~(\ref{eq:matrix4}),
and rate $\AlphadotStruct$ is used in the first row
to determine the stress rate $\Sigmadot$
(Eq.~\ref{eq:macrodstress}).
\par
We now use probe results to examine the directional nature of
dissipation in the macro-scale context.
The dissipation that is due to changes in variable
$\AlphaStruct$ is the product $\ChiStruct\cdot\AlphadotStruct$
(Eq.~\ref{eq:dissipform}),
and this dissipation is distinct from the much smaller dissipation
produced by changes in the fabric variable $\Alphafab$.
The stress-like $\ChiStruct$~--- the derivative of energy
$\psi$ with respect to $\AlphaStruct$~---
is found by combining Eqs.~(\ref{eq:psimacro1}), (\ref{eq:psi1a})
and~(\ref{eq:psi2macro}):
\begin{equation}\label{eq:xistrmacro}
  \ChiStruct =
  -\Nabla_{\AlphaStruct}[\Xibold]\,\psi
  =
  -\frac{\partial_{\boldsymbol{\xi}}\psi_{1}\big(\Varepsilon,
                           \{\Alphafab\},\AlphaStruct\big)}
        {\partial\AlphaStruct}
  -\frac{\partial_{\boldsymbol{\xi}}\psi_{2}\big(\AlphaStruct\big)}
        {\partial\AlphaStruct}
  =
  \Sigmabold
  - \big(\Sigmabold + \Nabla_{\AlphaStruct}[\Xibold] F_{2}
    \big)
  = - \Nabla_{\AlphaStruct}[\Xibold] F_{2}
\end{equation}
so that the macro-scale
$\ChiStruct$ is the directional derivative
of scalar function $F_{2}(\AlphaStruct)$.
We take the naive approach of treating $\ChiStruct$
as a G\^{a}teaux derivative, rather than a true gradient,
a choice that is confirmed by probe results.
\par
The nature of $\ChiStruct$
which shows the results of strain probes
conducted at two states during constant-$p$ loading:
during strain hardening and at the
critical state.
\begin{figure}
  \centering
  \includegraphics{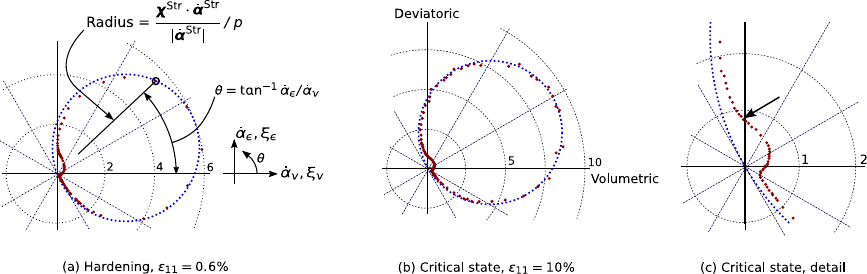}
  \caption{For the macro-scale model,
           plots of the dissipation stress
           $\ChiStruct$,
           assembled from the results of strain probes:
           (see Fig.~\ref{fig:Polarspars2}b for plotting method):
           (a)~during hardening, (b)~at the critical state, and
           (c)~detail at the critical state. 
           Angles $\theta$ correspond to the directions of
           rate $\AlphadotStruct$.
           Radial distances are the rate ratios
           $\ChiStruct\cdot\AlphadotStruct/\dot{\AlphaStruct}$
           which is equal to the directional derivative of
           $-F_{2}(\AlphaStruct)$ and to the magnitude of
           $\ChiStruct$.
           The arrow in~(c) identifies the probe that 
           continues the constant-$p$ loading at the critical state. 
           \label{fig:F2macro}}
\end{figure}
The plots are of the special type described in
Section~\ref{sec:plotting}, which allow direct visual assessment
of whether $\ChiStruct$ is a gradient (Fig.~\ref{fig:Polarspars2}b).
Each point represents a single probe carried out
with a directional strain increment $\Varepsilondot$,
in which the response~---
the rates of $\alpha^{\text{Str}}_{v}$, 
$\alpha^{\text{Str}}_{\epsilon}$,
and $F_{2}$~---
was measured.
The radial distance to each point is the quotient
$\ChiStruct\cdot\AlphadotStruct/|\AlphadotStruct|
=-\dot{F}_{2}/|\AlphadotStruct|$, 
and the angular position of the point denotes
the direction $\Xibold$ of $\AlphadotStruct$, as the angle
$\tan^{-1}(\dot{\alpha}^{\text{Str}}_{\epsilon}
/\dot{\alpha}^{\text{Str}}_{v})$.
All of the measured rates $\dot{F}_{2}$ were negative,
a result that is consistent with positive dissipation
$\ChiStruct\cdot\AlphadotStruct$
(see Eqs.~\ref{eq:dissipform} and~\ref{eq:chis}\textsubscript{1}).
Therefore, a radial distance is the magnitude of the
derivative
$\ChiStruct(\AlphaStruct;\Xibold)=-\Nabla_{\AlphaStruct}[\Xibold]F_{2}$
in the particular direction $\Xibold$ of $\AlphadotStruct$.
\par
Many of the points in Figs.~\ref{fig:F2macro}a
and~\ref{fig:F2macro}b
nearly lie on circles.
Although this approximate conformance to a circle would indicate
that the derivative $\ChiStruct$
is a true gradient $-\Nabla_{\AlphaStruct}F_{2}$,
this conclusion is misleading.
Fig.~\ref{fig:F2macro}c shows a detail of probe
points and their approximating circle near the origin of
the plot for the critical state.
The points in the detailed plot encompass probes
of $\Varepsilondot$ between the angles
$-125$\textdegree\ and $170$\textdegree
\ (refer to the previous Fig.~\ref{fig:macroalphaspars}c).
For example, the arrow in Fig.\ref{fig:F2macro}c
identifies the probe conducted in the 
direction of continued constant-$p$ loading.
The points in the detailed figure
do not conform to a circle passing through the plot's origin,
and a similar situation is found during hardening,
at the peak stress, and during post-peak softening.
We conclude that the dissipation stress $\ChiStruct$~---
the derivative of $-F_{2}(\AlphaStruct)$~---
is not a true gradient but must be
viewed as a G\^{a}teaux directional derivative
$-\Nabla_{\AlphaStruct}[\Xibold]F_{2}$.
\par
The points in Fig.~\ref{fig:F2macro}b
that lie close to the approximating circle and at the larger
radial distances (i.e., magnitudes of $|\ChiStruct|$)
are consistently of probes for which the 
strain direction $\Varepsilondot$ was in the range
$180$\textdegree--$270$\textdegree~---
strain probes that produce unloading in both
volumetric and deviatoric directions
(negative $\dot{v}$ and $\dot{\epsilon}$).
Points closer to the plot's origin (Fig.~\ref{fig:F2macro}c)
have smaller magnitudes $|\ChiStruct|$ and do not fit the circles.
The dissipation stresses $\ChiStruct$ of these points depart
most noticeably from a gradient $\Nabla_{\AlphaStruct}\psi$
and must be treated as a G\^{a}teaux derivative of $\psi$.
\par
We now complete the constitutive elements
by considering the scalar function $F_{2}(\AlphaStruct)$~---
a function that is directly computed with DEM results~---
and its relation to the macro-scale function
$\psi_{2}(\AlphaStruct)$ in
Eqs.~(\ref{eq:psimacro1}) and~(\ref{eq:psi2macro}).
In conventional elastoplasticity,
$\psi_{2}$ is associated with the back-stress of
kinematic hardening models \cite{Collins:1997a} and is referred
to as ``stored plastic energy'' or ``frozen elastic energy''
(see the works of Collins and coworkers
\cite{Collins:2002b,Collins:2005b}).
The back-stress $\boldsymbol{\rho}$ is the difference between
stress $\Sigmabold$ and the dissipation stress $\ChiStruct$,
or $\boldsymbol{\rho}=\Sigmabold-\ChiStruct$, and
is simply the directional derivative of $\psi_{2}(\AlphaStruct)$.
For macro-scale model, dissipation stress $\ChiStruct$ is the
G\^{a}teaux derivative shown in Fig.~\ref{fig:F2macro}.
The back-stress, therefore, is not a true gradient
but is a function of the direction $\Xibold$ of rate $\AlphadotStruct$.
\section{\normalsize Results in a micro-scale setting}\label{sec:microresults}
In the micro-scale approach,
we selected quantities $\Alphastruct$
and $\Chistruct$ that are aggregates of a material's
contact-level sliding rates and frictional forces,
and we hypothesize that $\Chistruct$ is the directional derivative
of energy $\psi$ with respect to $\Alphastruct$.
(in this section we use the lowercase ``str'' as superscripts).
Both quantities are measured in DEM simulations
and are scaled so that the product
$\Chistruct\cdot\Alphadotstruct$ matches the measured
dissipation rate
(see Eqs.~\ref{eq:chistr1}--\ref{eq:alphadis}
and \ref{sec:micro4}).
In this section,
we review these quantities during both monotonic loading
and series of strain probes.
\subsection{Micro-scale approach~--- derivative $\Chistruct$}
\begin{figure}
  \centering
  \includegraphics{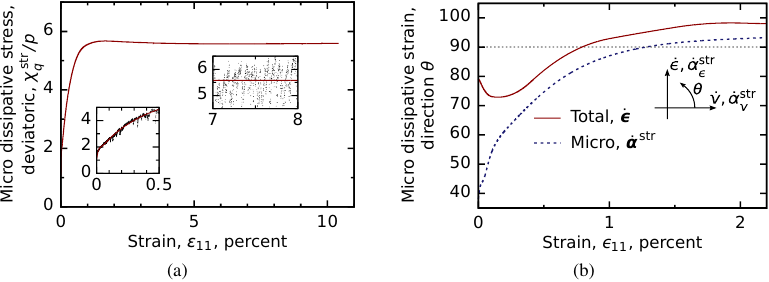}
  \caption{For micro-scale model, results during constant-$p$
           loading:
           (a)~deviatoric $q$ component 
           of dissipative stress $\chi^{\text{str}_{q}}$
           (Eq.~\ref{eq:chistr1}),
           divided by the bulk mean stress $p$, and
           (b)~direction of the micro dissipative rate
           $\Alphadotstruct$, compared with the strain rate
           $\Varepsilondot$
           (Eqs.~\ref{eq:alphadotpre}--\ref{eq:alphadis}).
           Angles $\theta$ greater than 90\textdegree\ are dilative;
           angles less that 90\textdegree\ are contractive.
           The peak stress occurred at 1.8\% strain.
           \label{fig:ChiStrMicro4QVsEps11}}
\end{figure}

Fig.~\ref{fig:ChiStrMicro4QVsEps11}a shows the evolution of the
deviatoric component $\chi^{\text{str}}_{q}$
of the dissipative stress
$\Chistruct$ (i.e.,
the stress
attributed to the magnitudes of tangential forces at sliding
contacts)
during triaxial constant-$p$ loading.
As was done with the full deviatoric stress $q$
in Fig.~\ref{fig:constp}b,
the stress-like $\chi^{\text{str}}_{q}$ is normalized
by dividing by the full mean stress $p$.
The deviatoric ``$q$'' component
is shown as a smoothed spline fit of the data,
and representative scatter is shown in the two insets.
The scatter is attributed to the changing subset
of contacts that are sliding,
as some contacts alternate between a sliding and non-sliding condition,
a result of their particles' sometimes tenuous internal arrangements
(see \cite{Kuhn:2025a}).
\par
Comparing the stresses $\chi^{\text{str}}_{q}$ and $q$ in
Figs.\ref{fig:ChiStrMicro4QVsEps11}a and~\ref{fig:constp}b,
the values of $\chi^{\text{str}}_{q}$,
which are restricted to the tangential forces
at the sliding contacts,
are consistently larger than stress $q$,
even though the subset of sliding contacts
accounts for only 10\%--35\% of all contacts.
The  $\chi^{\text{str}}_{q}$, however, is computed
from the \emph{magnitudes} of the tangential components of force,
$|\mathbf{f}^{\text{t}}|$;
whereas $q$ derives from the directed forces $\mathbf{f}$.
The latter set is highly disordered, so that directed forces
cancel each other, moderating their contribution to $q$.
\par
The non-zero dissipative
stress $\chi^{\text{str}}_{q}$ upon the start
of loading deserves explanation
(see inset in Fig.~\ref{fig:ChiStrMicro4QVsEps11}a).
Frictional dissipation was seen in Fig.~\ref{fig:dissip1}a
to begin at the start of loading,
and in the probe plot of Fig.~\ref{fig:PolarDis}a
dissipation is seen to occur in all directions of
initial loading (solid black line in Fig.~\ref{fig:PolarDis}a).
This evidence suggests that before loading many contacts
were poised to slide, even though the prior
isotropic consolidation produced zero deviatoric stress $q$.
The value of $\chi^{\text{str}}_{q}$
was actually zero at $\varepsilon_{11}=0$,
and the non-zero $\chi^{\text{str}}_{q}$ shown in the
inset of Fig.~\ref{fig:ChiStrMicro4QVsEps11}a
is recorded at the tiny strain $\varepsilon_{11}=0.0001\%$.
These results demonstrate that a group of contacts is
initially ready to slide,
but that sliding and dissipation
for a subset of this group does not
occur until loading begins, thus immediately
producing a non-zero $\chi^{\text{str}}_{q}$.
Moreover,
this initial sliding and dissipation occurs
regardless of the direction of loading
(again, Fig.~\ref{fig:PolarDis}a),
although sliding occurs
among different subsets of the poised group for
different loading directions.
\par
Figure~\ref{fig:ChiStrMicro4QVsEps11}b
shows the evolving direction of the strain-like
$\Alphadotstruct$ during early constant-$p$ loading.
The direction of $\Alphadotstruct$ is computed from the
directions of sliding among the contacts
(the same direction as the precursor strain $\Alphadotpre$
in Eq.~\ref{eq:alphadotpre});
whereas the magnitude of $\Alphadotstruct$ is computed from the
measured dissipation, so that $|\Alphadotstruct$| is
energy-consistent with its fluctuating counterpart $\Chistruct$
(see Eq.~\ref{eq:alphadis}).
The dissipative $\Alphadotstruct$ is seen to be
consistently more compressive than strain $\Varepsilondot$.
\subsection{Micro-scale approach~--- strain probes}
We now consider the results of strain probes in the
micro-scale setting of Section~\ref{sec:micro2},
Values of $\Chistruct$ and $\Alphadotstruct$ were measured
in these probes, with the purpose of
appraising the differentiability of energy $\psi$
with respect to micro-scale $\Alphastruct$.
In the next section, the same probes are used to determine
elements of a constitutive model.
\par
\begin{figure}
  \centering
  \includegraphics{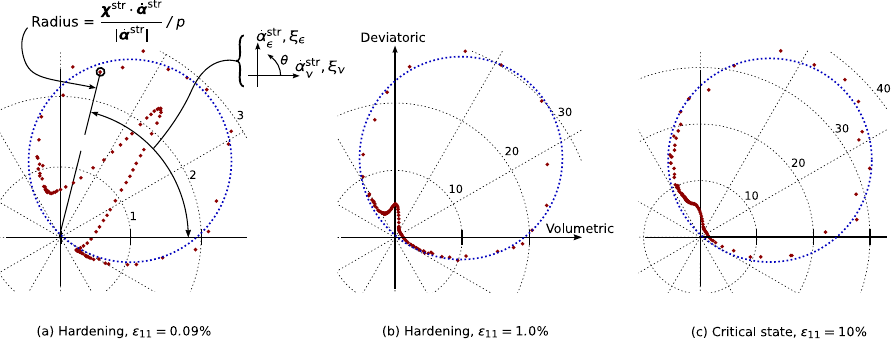}
  \caption{For the micro-scale model, plots of probe results:
           quotient of the scalar dissipation rate
           $\Chistruct\cdot\Alphadotstruct$
           due to structural variable $\Alphastruct$
           and the rate magnitude $|\Alphadotstruct|$
           (see Fig.~\ref{fig:Polarspars2}c for plotting method).
           For the micro-scale model,
           $\Chistruct$ and $\Alphastruct$ are
           derived from micro-scale contact movements and forces.
           Three states of prior constant-$p$ monotonic
           loading are shown: (a)~hardening domain,
           (b)~peak deviator stress, and
           (c)~critical state.
           Angles $\theta$ correspond to the directions
           of the micro-scale rates $\Alphadotstruct$
           (i.e., of the unit vectors $\Xibold$).
           Radii are the rate ratios
           $\Chistruct\cdot\Alphadotstruct/|\Alphadotstruct|$,
           thus are the magnitudes $|\Chistruct|$.
           Results are normalized by dividing by mean stress $p$.
           The dashed blue circles are approximations of the data.
           \label{fig:Polar_dF2_dalpha_micro4}}
\end{figure}

Fig.~\ref{fig:Polar_dF2_dalpha_micro4} shows polar
plots of the dissipation rate
$\Chistruct\cdot\Alphadotstruct$ attributed to
the structure variable $\Alphastruct$
for strain probes at three states during
constant-$p$ loading.
This part of dissipation is computed with
Eq.~(\ref{eq:chis}\textsubscript{2})
and is separate from a much smaller dissipation
produced by changes in the fabric variable $\Alphafab$.
In accord with the plotting methods of Section~\ref{sec:plotting},
the radii in Fig.~\ref{fig:Polar_dF2_dalpha_micro4}
are dissipation rates
$\Chistruct\cdot\Alphadotstruct$ divided by
the rate magnitudes $|\Alphadotstruct|$;
the direction angles $\theta$ are of the rates $\Alphadotstruct$
that were induced by probe strains $\Varepsilondot$.
(Note that strains $\Varepsilon$ can be controlled,
unlike the internal variable $\Alphastruct$, so that
$\Alphadotstruct$ is simply a measured response to the imposed
$\Varepsilondot$.)
In the two-dimensional setting of triaxial probes,
direction $\theta$ is that of $\Alphadotstruct$ and of its
unit vector $\Xibold$.
\par
The figure, therefore, shows the magnitude of the stress-like derivative
$\Chistruct(\Varepsilon,\Alphastruct ; \Xibold)$ as
a function of direction $\Xibold$.
The two parameters $\Varepsilon$ and $\Alphastruct$ were
established by the prior constant-$p$ loading, and only
direction $\Xibold$ varied in the probes.
If derivative $|\Chistruct|$ is a true gradient~---
if $\psi$ is differentiable with respect to $\Alphastruct$~---
then the results would conform to a circle passing through the origin
(as in Fig.~\ref{fig:Polarspars2}b).
The dashed circles in Fig.~\ref{fig:Polar_dF2_dalpha_micro4}
do, in fact, approximate many of the data points
(these points are for strain probes that are
roughly aligned with the prior constant-$p$ loading),
and diameters and orientations of the circles are the
magnitudes and orientations of an approximating gradient
$\Nabla_{\alpha^{\text{str}}}\psi$.
However,
the data deviates from these approximating circles,
particularly for unloading strains,
This result
demonstrates that a G\^{a}teaux derivative of $\psi$
is necessary for evaluating dissipation with the micro-scale model.
\subsection{Micro-scale approach~--- constitutive elements}
Three incremental constitutive forms are presented in
Section~\ref{sec:constituteforms}.
Although all three can describe the same behavior, the
Form~2 of Eq.~(\ref{eq:matrix4})
is more directly constructed from DEM data
when the structure variable $\Alphastruct$ corresponds to
the micro-scale sliding rates of contacts.
We describe how the components of Form~2 are gathered from this data.
The second row of Eq.~(\ref{eq:matrix4}) is a set of $n$
constraints $\{\boldsymbol{\phi}\}$
that link the strain rates $\Varepsilondot$
and the rates $\Alphadotset$.
For triaxial conditions,
the latter rates are reduced to the rates of two fabric
variables (Sections~\ref{sec:moduli2}--\ref{sec:moduli3})
and the two components,
deviatoric and volumetric, of $\Alphadotstruct$
(Section~\ref{sec:micro2}).
The DEM results provide this constitutive relationship
between $\Varepsilondot$ and $\Alphadotstruct$.
The contribution of $\Alphadotstruct$ to the stress rate $\Sigmadot$~---
the term $\mathbf{H}^{\Varepsilon\boldsymbol{\alpha}}$
in the first row of Eq.~(\ref{eq:matrix4})~---
is also provided by DEM results and
discussed below, thus fulfilling
the material's constitutive elements.
\par
\begin{figure}
  \centering
  \includegraphics{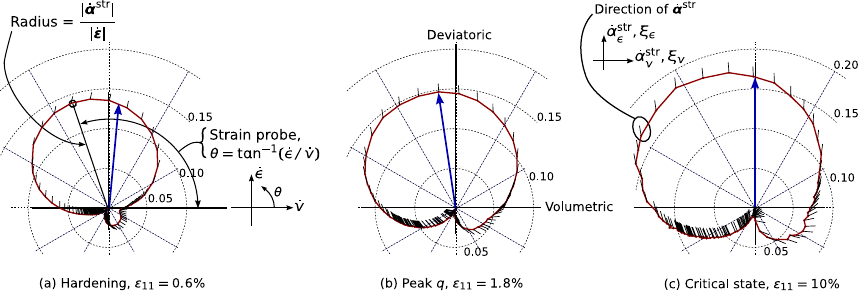}
  \caption{For the micro-scale model,
           plots of probe results,
           giving the relationship between the vector rates
           $\Varepsilondot$ and $\Alphadotstruct$
           (see Fig.~\ref{fig:Polarspars2}c for plotting method).
           Three states of the prior constant-$p$ monotonic
           loading are shown: (a)~hardening domain,
           (b)~peak deviator stress, and
           (c)~critical state.
           Angles $\theta$ correspond to strain directions
           $\Varepsilondot$,
           and radial distances to the red lines give the rate ratio
           $|\Alphadotstruct|/|\Varepsilondot|$ for these
           strain directions.
           The short spars point in the direction of
           rates $\Alphadotstruct$.
           The blue arrows are the direction of $\Varepsilondot$
           during the prior constant-$p$ loading.
           See Section~\ref{sec:plotting}
           and Fig.~\ref{fig:Polarspars2}c for
           context on the plotting method.
           \label{fig:PolarSparsEpsDalphaMicro4}}
\end{figure}

Fig.~\ref{fig:PolarSparsEpsDalphaMicro4} shows the relationship
between rates $\Varepsilondot$ and $\Alphadotstruct$
for three states along the constant-$p$ load path.
The figure encapsulates the
second row of constitutive Form~2 at each state,
with each data point corresponding to a single strain probe
$\Varepsilondot$.
The polar plots are of the type described in Section~\ref{sec:plotting},
with radii being a ratio of rates
$|\Alphadotstruct|/|\Varepsilondot|$, and angle $\theta$
giving the direction of strain rate $\Varepsilondot$.
The small spars that emanate from data points are
in the directions of $\Alphadotstruct$ (i.e., $\Xibold$)
that are induced by $\Varepsilondot$.
As a reference direction, the constant-$p$ loading
that preceded the probes produced rates $\Varepsilondot$ with
orientations of 73\textdegree--$\,$98\textdegree.
\par
During hardening (Fig.~\ref{fig:PolarSparsEpsDalphaMicro4}a),
the polar plot is approximately composed of two circles:
a larger upper circle for probes with deviatoric
rates $\dot{\epsilon}>0$ (roughly corresponding to the prior
constant-$p$ loading),
and a smaller circle for unloading rates.
Spars that emanate from the upper circle are roughly aligned in a
common direction having a predominantly positive deviator
rate $\dot{\alpha}_{\epsilon}>0$;
spars on the smaller circle point in directions of
a negative rate $\dot{\alpha}_{\epsilon}<0$.
These results indicate that, at low strains,
the rate $\Alphadotstruct$ is described as the sum of
two dyads multiplied by $\Varepsilondot$
(see the discussion of Fig.~\ref{fig:Polarspars2}c).
For example,
at strain $\varepsilon_{11}=0.6\%$ 
(Fig.~\ref{fig:PolarSparsEpsDalphaMicro4}a),
rate $\Alphadotstruct$ is approximated as the sum of
two mechanisms
\begin{equation}\label{ref:twomech}
  \Alphadotstruct =
  \sum_{\text{i}=1,2}
  \gamma^{(\text{i})}\mathbf{h}^{(\text{i})}\langle
  \mathbf{f}^{(\text{i})}\cdot\Varepsilondot\rangle
\end{equation}
%
Reading directly from Fig.~\ref{fig:PolarSparsEpsDalphaMicro4}a,
the diameters $\gamma^{(\text{i})}$ of the larger
and smaller circles are 0.14 and~0.024;
the unit directions $\mathbf{g}^{(\text{i})}$ of the
circles' orientations are
at 110\textdegree\ and~300\textdegree;
and the unit spar directions $\mathbf{h}^{(\text{i})}$ are
at 85\textdegree\ and~330\textdegree.
\par
At larger strains, the relationship between the rates
$\Varepsilondot$ and $\Alphadotstruct$ is more irregular
(Figs.~\ref{fig:PolarSparsEpsDalphaMicro4}b,c).
The variable's rate $\Alphadotstruct$,
illustrated with spars,
is seen to transition through almost all directions $\Xibold$
as the direction of $\Varepsilondot$ is rotated.
Moreover,
the magnitude of rate $|\Alphadotstruct|$, given by the plot's
radii, is no longer described by a pair of circles, but
varies in an irregular manner with the direction of $\Varepsilondot$.
The plot shows that
relationship between $\Varepsilondot$ and $\Alphadotstruct$
can not be reduced to a simple form such as Eq.~(\ref{ref:twomech})
when loading reaches the peak stress and beyond.
\par
Comparing the results for the macro- and micro-scale models,
Figs.~\ref{fig:macroalphaspars}
and~\ref{fig:PolarSparsEpsDalphaMicro4} are quite similar in their
shapes and in their spar directions.
The structure variables
$\AlphaStruct$ and $\Alphastruct$ of the two models, which were
developed from different perspectives~---
the macro-scale irreversible strain and a micro-scale
measure of contact sliding~---
seem to share a common origin.
\par
\begin{figure}
  \centering
  \includegraphics{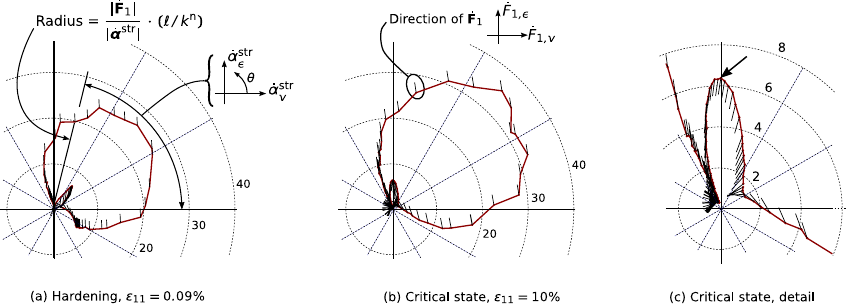}
  \caption{For the micro-scale model, plots of probe results,
           giving the relationship between the vector rates
           $\Alphadotstruct$ and $\dot{\mathbf{F}}_{1}$
           (see Fig.~\ref{fig:Polarspars2}c for plotting method).
           Three states of the prior constant-$p$ monotonic
           loading are shown:
           (a)~hardening at 0.09\% strain,
           (b)~at the critical state, and
           (c)~a detail at the critical state.
           Radii are normalized by multiplying by the quotient
           of particle size $\ell$ and contact stiffness $k^{\text{n}}$.
           Angles $\theta$ correspond to directions of $\Alphadotstruct$,
           and radial distances to the red lines give the rate ratio
           $|\dot{\mathbf{F}}_{1}|/|\Alphadotstruct|$ for these
           directions.
           The spars point in the direction of
           rates $\dot{\mathbf{F}}_{1}$, normalized with quotient
           $\ell/k^{\text{n}}$ (Table~\ref{table:assembly}).
           The arrow in~(c) identifies the probe that 
           continues the constant-$p$ loading at the critical state.
           See Section~\ref{sec:plotting} 
           and Fig.~\ref{fig:Polarspars2}c for context on these plots.
           \label{fig:PolarSparDF1Micro4}}
\end{figure}

As the final constitutive piece of the micro-scale model,
Fig.~\ref{fig:PolarSparDF1Micro4} shows the relationship between
rate $\dot{\mathbf{F}}_{1}$ and the micro-scale structural rate
$\Alphadotstruct$.
The function $\mathbf{F}_{1}(\Alphastruct)$ is
an additive and irreversible contribution to stress $\Sigmabold$
(Eqs.~\ref{eq:sigmagen} and~\ref{eq:variables}),
and radial distances in the plot are magnitudes of
the derivative of $\mathbf{F}_{1}(\Alphastruct)$ in directions
$\Xibold$ of $\Alphadotstruct$.
The spars show the directions of rates $\dot{\mathbf{F}}_{1}$.
The figure, therefore, depicts the stress modulus
$\mathbf{H}^{\Varepsilon\boldsymbol{\alpha}}$
as a function of
$\Xibold$,
noting that $\mathbf{H}^{\Varepsilon\boldsymbol{\alpha}}$ is the
stiffness that appears in the upper right corner
of the constitutive matrices of all three forms in
Section~\ref{sec:constituteforms}
(also, Eq.~\ref{eq:defHea}\textsubscript{1}).
The total stress rate $\dot{\Sigmabold}$ is the sum of the reversible
rate $\Erev\cdot\Varepsilondot$ and
an irreversible rate $\dot{\mathbf{F}}_{1}$ equal to the product
$\mathbf{H}^{\Varepsilon\boldsymbol{\alpha}}\cdot\Alphadotstruct$.
\par
Figures.~\ref{fig:PolarSparDF1Micro4}a and~b correspond to
the hardening regime and to the critical state.
Each plot is composed roughly of two parts:
a small elongated part near the origin,
and a larger and more rotund part.
The first part applies to probes with direction $\dot{\epsilon}>0$
(i.e., approximately in directions of continued deviatoric loading).
Its elongated non-circular shape
clearly demonstrates that the directional
derivative of $\mathbf{F}_{1}$ with respect to $\Alphastruct$
is not a true gradient, so that modulus
$\mathbf{H}^{\Varepsilon\boldsymbol{\alpha}}$
must be treated as a generalized derivative, particularly
during continued deviatoric loading.
The downward direction of the spars (a negative $\dot{F}_{1,\epsilon}$,
Fig.~\ref{fig:PolarSparDF1Micro4}c)
means that $\dot{F}_{1,\epsilon}$,
the irreversible contribution to stress rate
$\dot{\Sigmabold}$, counteracts the
reversible contribution of strain $\Varepsilondot$.
\par
The more rotund second parts in
Figs.~\ref{fig:PolarSparDF1Micro4}a and~b correspond
to unloading probes ($\dot{\epsilon}<0$),
and with these strain directions,
the directional spars of $\dot{\mathbf{F}}_{1}$ are
roughly pointing upward in a common direction
(in the direction of increasing deviatoric stress $\dot{q}$).
Again, the irreversible contribution to stress rate
$\dot{\Sigmabold}$ counteracts the reversible contribution.
The rotund, nearly circular shape during unloading indicates that
when $\dot{\epsilon}<0$ the modulus
$\mathbf{H}^{\Varepsilon\boldsymbol{\alpha}}$ approximately
represents a true gradient of stress with respect to
$\Alphastruct$.
As such, rate $\dot{\mathbf{F}}_{1}$
can be approximated as
$\dot{\mathbf{F}}_{1}
\approx \mathbf{g}\langle\mathbf{f}\cdot\Alphadotstruct\rangle$,
where unit vector $\mathbf{g}$ has an upward direction of
about 95\textdegree, unit vector $\mathbf{f}$ is the
orientation of the circles (about 45\textdegree),
and $\gamma$ is a circles' diameter.
\section{\normalsize Discussion}\label{sec:conclusions}
We have proposed that energy and dissipation in dense granular
materials are controlled by two sets of internal variables:
a set of fabric variables that governs the reversible stiffness
(i.e. stiffness in the absence of sliding) and a set
of structure variables
that is associated with the sliding among particles.
With the former,
reversible stiffness depends upon
the number, stiffnesses, and orientations of the contacts,
so that the relevant fabric variables $\{\Alphafab\}$
are a short list of these micro-scale factors
(see Eq.~\ref{eq:Erevdep}).
Although we did not investigate reasons for changes
in the fabric variables
(i.e, in the rates $\{\Alphadotfab\}$),
we found that a small dissipation is attributed to the coupling
of modulus $\Erev$ and the changing fabric,
primarily during hardening
(Fig.~\ref{fig:dissip1}).
This dissipation due to $\{\Alphadotfab\}$
suggests that changes in the fabric variables
are only brought about by a rearrangement of particles
that requires sliding among the particles,
a sliding that is registered with the structure rate $\Alphadotstruct$.
That is, the two sets of variables~--- fabric and structure~---
are likely coupled, a matter that was not addressed
with the simulations and requires further inquiry.
\par
Apart from the small portion of dissipation that is attributed to
fabric changes,
the simulations demonstrate
the ubiquity of sliding and dissipation,
as chronicled with the structure variable $\Alphastruct$.
We found that contact sliding can occur at
the very start of deviatoric loading, even when the
preceding consolidation is entirely isotropic.
Whenever any subsequent monotonic loading is halted,
sliding occurs in all probe directions, whether loading or unloading,
thus negating any purely elastic domain in strain-space.
Among these confounding results,
we found that some of the
contacts that were sliding during monotonic loading
continue to slide during an abrupt reversal of the loading,
meaning that contact movements are not reversed when
loading is reversed.
Indeed, the directions of contact movements during loading
and unloading are nearly uncorrelated.
These irregular, even surprising, results highlight
the pervasiveness of dissipation during deformation
and indicate a complex evolution of the energy function $\psi$.
\par
We have taken advantage of the ability to measure
energy, dissipation, stress, and strain in DEM simulations.
By measuring these essential thermomechanic quantities, the
functions $\mathbf{F}_{1}(\Alphastruct)$ and
$F_{2}(\Alphastruct)$ are calculated along a load path.
The two functions then allow candidate variables
$\Alphastruct$ to be quantified and tested.
For both macro- and micro-scale models,
the results show, unfortunately, that neither $\mathbf{F}_{1}$
nor $F_{2}$ is truly a function of $\Alphastruct$.
In the probe results
of Figs.~\ref{fig:F2macro} and~\ref{fig:PolarSparDF1Micro4},
each probe began from same $\Alphastruct$, but
a single rate $\Alphadotstruct$
could map to \emph{more than one rate} of $\mathbf{F}_{1}$
and $F_{2}$: a ray drawn from the origin can meet two points
along the locus of points $\dot{\mathbf{F}}_{1}$
and $\dot{F}_{2}$.
The two intersection points correspond to
different directions of the probe strain $\Varepsilondot$.
This result suggests that the rates $\dot{\mathbf{F}}_{1}$
and $\dot{F}_{2}$ are functions of the direction of
$\Varepsilondot$ as well as that of $\Alphadotstruct$.
\par
Simulations also show that $\psi$
is not truly differentiable with respect to
the internal variables $\Alphaset$,
in particular, with respect to the structure variable
$\Alphastruct$.
Whether this variable is viewed as the irreversible strain
$\AlphaStruct$
or as a micro-scale measure of sliding $\Alphastruct$,
the complementary stress $\Chistruct$ must be treated as
a general G\^{a}teaux derivative of $\psi$.
This result is candidly
exposed by our special plotting of the derivatives of
$\mathbf{F}_{1}$ and $F_{2}$
in Figs.~\ref{fig:F2macro} and~\ref{fig:Polar_dF2_dalpha_micro4}.
The odd, non-circular shapes of the plots indicate a
complex landscape of $\psi$,
with $\psi$ having a local topography in which
a local gradient does not apply.
The reversible moduli $\Erev$, however, are found to be
sufficiently regular
to treat the derivative of $\psi$ with respect to the fabric variables
$\{\Alphafab\}$ as a gradient.
In regard to the constitutive forms that were discussed
in Section~\ref{sec:constituteforms},
the Form~3 in Eq.~(\ref{eq:matrix3}) is most commonly used in
elastoplastic models when all derivatives are simple gradients
that are independent of direction $\Alphadotstruct$.
Because more general G\^{a}teaux derivatives apply,
the general constitutive Form~2 in Eq.~(\ref{eq:matrix4})
is more appropriate,
as direct use is made
of the relationship between rates $\Varepsilondot$ and $\AlphadotStruct$.
\par
We now consider whether energy $\psi$
exhibits path independence
when its arguments are limited to strain $\Varepsilon$
and the variables $\{\Alphafab\}$ and $\Alphastruct$:
whether starting at a state having energy
$\psi_{\text{o}}$ with the state variables
$\{\Varepsilon_{\text{o}}, \{\Alphafab_{\text{o}}\},
\Alphastruct_{\text{o}}\}$,
all paths through state-space that end at the original
state variables (i.e., a closed path) also end with the
original energy $\psi_{\text{o}}$.
That is, path independence requires a vanishing integral,
$\oint d\psi = 0$, for any closed path
when the differential 1-form $d\psi$ is given by
Eq.~(\ref{eq:balance}).
The gradient theorem holds that
for a function $f(\{\mathbf{x}\})$ of variables $\{\mathbf{x}\}$,
path independence is assured if $f$ is differentiable
(i.e., with a gradient $\Nabla_{\{\mathbf{x}\}} f$),
in which case $\oint\Nabla f (\{\mathbf{x}\})\cdot d\{\mathbf{x}\}=0$
for all closed paths.
Differentiability, however,
is sufficient but not necessary for path independence.
(As a counter-example, the earth's topography is not
differentiable, noting its pointed ridges and abrupt bluffs,
but the difference in elevation between two points
can always be reckoned by integrating slope and distance
along a footpath.)
\par
The issue of path independence is whether
an energy function
$\psi(\Varepsilon,\{\Alphafab\},\Alphastruct)$ that
is not differentiable with respect to $\Alphastruct$,
except as a G\^{a}teaux derivative,
will return to the same $\psi_{\text{o}}$ after a
closed path is traversed in state-space, returning to the original
state variables.
For the macro-scale model, in which $\AlphaStruct$ is the
cumulative irreversible strain,
path independence was not tested in the paper and remains an
open question.
With the paper's micro-scale model, however,
we make the following argument of \emph{approximate} path independence.
Energy $\psi$ is, in essence, the expected value of the squared
magnitudes of contact forces (Eq.~\ref{eq:Helm}).
The relevant question is whether it is
possible, then, for two particle arrangements
to share the same state variables but to have different
$\psi$ values?
In other words, are the state variables
$\{\Varepsilon_{\text{o}}, \{\Alphafab_{\text{o}}\},
\Alphastruct_{\text{o}}\}$ sufficient to yield a unique
expected value of the squared contact forces?
\par
Answering these question is more manageable by considering
a complementary set of state variables,
noting that
the state variables of $\psi$ can be classified as two types:
(1)~active variables associated with a change in condition and
measured relative to some reference state
(e.g., the strain $\Varepsilon$
and the sliding-induced strain-like $\Alphastruct$),
and (2)~passive variables determined at a current condition or state
(e.g., stress $\Sigmabold$, the fabric variables
$\{\Alphafab\}$ in Eq.~\ref{eq:Erevdep},
and the stress-like $\Chistruct$ in Eq.~\ref{eq:chistr1}).
Through Legendre transformations \cite{Collins:1997a},
one can define a free energy $\Legendre$ whose state variables
are all passive,
\begin{equation}\label{eq:Legendre}
  \Legendre\big(\Sigmabold,\{\Alphafab\},\Chistruct\big)
  \;=\;
  \psi\big(\Varepsilon,\{\Alphafab\},\Alphastruct\big)
  \;-\;\Sigmabold\cdot\Varepsilon
  \;+\;\Chistruct\cdot\Alphastruct
\end{equation}
thus retaining the fabric variables $\{\Alphafab\}$
(contact density, contact anisotropy, etc.) but shifting to
quantities $\Sigmabold$ and $\Chistruct$,
which are aggregate measures of the current contact forces.
\par
From a particle-scale perspective, the variables
$\{\Sigmabold, \{\Alphafab_{\text{o}}\},
\Chistruct\}$ of the transformed energy $\Legendre$
greatly constrain the possible values of $\psi$.
These variables are the expected value of contact force
(the mean stress $p$, Eq.~\ref{eq:stress} and~\cite{Cundall:1983a});
the expected force orientation
(the deviatoric part of stress $\Sigmabold$, Eq.~\ref{eq:stress});
the contact density ($\varrhoc$);
the expected contact orientation
($\boldsymbol{\beta}^{\text{fab}}$, Eq.~\ref{eq:defs1}\textsubscript{2});
and the expected magnitude and orientation
of the sliding contact forces ($\Chistruct$, Eq.~\ref{eq:chistr1}).
This list of expected values greatly restrict
the possible values of $\psi$
(again, the expectation of squared contact forces).
For this reason, we suggest that $\psi$ is
\emph{almost uniquely determined} by the chosen list of
micro-scale variables.
\par
Finally, we note that these thermomechanic principles
were explored with simulations of
particles having a single non-convex shape (Fig.~\ref{fig:constp}a),
a shape that was chosen to simulate the behavior of a
particular granular soil (Nevada Sand).
The simulation results are also in qualitative
agreement with those of other investigators
using sphere assemblies and other
particle shapes (many cited herein).
Closer quantitative conformance with a target material,
however, requires simulations that share the particle shapes,
bulk density, initial fabric, and contact characteristics
of the intended material.
\section*{\normalsize Data availability}
The data used in the paper is available as a compressed file
that contains the source code of the DEMPLA/OVAL software
for the DEM simulations,
the output of the simulation runs,
Octave/Matlab files for processing the raw output,
files of the processed data, and plotting files
used in creating the paper's figures:
\texttt{https://faculty.up.edu/kuhn/misc/misc.html}.
\appendix
\section*{\normalsize Appendices}
\section{\normalsize Quantities measured with DEM}%
\label{sec:measuredem}
Table~\ref{table:measure} lists thermomechanic
quantities that are
measurable with DEM.
Most straightforward,
the average strain $\Varepsilon$ is measured with the
displacements of an assembly's boundary particles
(the displacement of periodic boundaries in
the paper's simulations).
The average
stress $\Sigmabold$ is directly
computed from the contact forces among particles,
with the Love--Weber equation:
\begin{equation}\label{eq:stress}
  \sigma_{ij} =
  \varrhoc\:
  \Ec(f_{i}l_{j})
  \quad\text{or}\quad
  \Sigmabold =
  \varrhoc\:
  \Ec(\mathbf{f}\otimes\mathbf{l})
\end{equation}
where $\Ec(\cdot)$ is the expected (mean) value of
a set of quantities,
with each quantity belonging to
one of the assembly's $M^{\prime}$ contacts.
In this case, $E(f_{i}l_{j})$ is the expected value of
the component-wise scalar products $f_{i}l_{j}$ of
contact forces $\mathbf{f}$ and branch vectors
$\mathbf{l}$
(a branch vector connects the centers of two contacting particles).
As an alternative,
tensor $E(\mathbf{f}\otimes\mathbf{l})$
is the expectation of contact dyads
$\mathbf{f}\otimes\mathbf{l}$.
Scalar $\varrhoc=M^{\prime}/V$ is the contact density~---
contacts per volume.
This accounting of stress only considers
stress within the grain bodies and ignores
stress within inter-granular voids
(i.e., that of the pore fluid), so that the
$\Sigmabold$ considered herein is an
effective stress.
Because all quantities in the equation are measured
in the current, displaced configuration,
$\Sigmabold$ is the Cauchy stress.
\par
The Helmholtz energy $\psi$ of a granular region
is the volume-average of elastic energies within a
region's particles,
and with DEM models, elastic energy is idealized
as being held within an assembly's contact springs.
The linear--frictional contact model used herein
consists of linear normal springs and tangential linear spring
in series with frictional sliders, so that
the Helmholtz energy $\psi$ is simply the
product of contact density $\varrhoc$
and the expected value of energy in individual contacts,
\begin{gather}\label{eq:Helm}
  \psi =
  \varrhoc\:\Ec
  \Big(
  (f^{\text{n}})^{2}/(2k^{\text{n}})
  + 
  |\Ft|^{2}/(2k^{\text{t}})
  \Big)
  \\
  f^{\text{n}} =f_{i}n_{i},
  \quad
  f_{i}^{\,\text{t}} =
  f_{i} - f^{\,\text{n}} n_{i}
\end{gather}
where $k^{\text{n}}$ and $k^{\text{t}}$ are the
contacts' spring stiffnesses;
$f^{\text{n}}$ and $\Ft$ are the
normal and tangential forces;
and $\mathbf{n}$ are the normal unit vectors.
\par
If the DEM model is one with resilient, non-breaking particles
in which tangential sliding is the sole form of
dissipation
(e.g., materials without grain fracture or plasticity,
without viscosity in the particle movements, etc.),
the dissipation rate
$\Chiset\cdot\Alphadotset$ in
Eqs.~(\ref{eq:balance})--(\ref{eq:ineq})
is the volume-average frictional dissipation
at sliding contacts.
With a linear--frictional contact model,
a contact's slip velocity
$\Deltaslip$
is tangential and
occurs in the direction of the tangential contact force
$\Ft$ \cite{Kuhn:2020b},
and a contact's dissipation rate
the product of its
tangential force and slip velocity \cite{Kruyt:2006a}:
\begin{equation}\label{eq:DEMdissip}
  \Chiset\cdot\Alphadotset
  =
  \varrhocsl\:\Esl
  \big(
  f_{i}^{\,\text{t}}\deltaslip_{i}
  \big)
\end{equation}
in which the expected (mean) value $\Esl(\cdot)$ is that
of the set of \emph{sliding contacts},
and the contact density $\varrhocsl$ is the number of
sliding contacts per volume.
\par
The reversible moduli
$\Erev=
\partial^{2}\psi/\partial\boldsymbol{\varepsilon}^2$
are measured by imposing
small increments of deformation (strain probes)
while preventing dissipative movements at the particles' contacts:
preventing frictional slip
but allowing elastic contact movements.
Slip was prevented by assigning a large friction coefficient
to the contacts for the duration of the probe.
By conducting multiple probes in different strain
directions,
one can develop the full stiffness tensor
$\Erev$
(see \cite{AlonsoMarroquin:2005b,Calvetti:2003a,Kuhn:2018c}
for similar methods).
\section{\normalsize Estimates of $\Alphastruct$ and $\Chistruct$}%
\label{sec:micro4}
Section~\ref{sec:micro2}
provides three criteria
for selecting $\Alphastruct$ and $\Chistruct$, such that the
actual dissipation, the scalar $\varrhocsl\Esl(\Fst_{i}\deltaslip_{i})$,
derives from the product
$\Chistruct\cdot\Alphadotstruct$ (see Eq.~\ref{eq:DEMdissip}).
We adopt the tensors given in
Eqs.~(\ref{eq:chistr1})--(\ref{eq:alphadis}), but only after
several failed candidates~---
specifically, candidates that did not
consistently satisfy the criterion
$\Chistruct\cdot\Alphadotstruct\ge 0$~---
due to the candidates' $\Alphadotstruct$ and $\Chistruct$
occasionally being counter-aligned.
Before deriving
Eqs.~(\ref{eq:chistr1})--(\ref{eq:alphadis}),
we begin with one failed attempt
so that, perhaps, others can avoid similar pitfalls.
\par
As an attempt at a stress-like $\Chistruct$ based
on the frictional contact forces $\Ft$ of sliding contacts,
we make the seemingly obvious choice of replacing the
full stress
$\sigma_{ij}=\varrhoc\Ec(f_{i}l_{j})$
in Eq.~(\ref{eq:stress}) with a stress-like counterpart
$\boldsymbol{\chi}^{\text{str,?}}$
(the ``?'' denotes its tentative status),
\begin{equation}\label{eq:chiq}
  \chi^{\text{str,?}}_{ij}
  =
  \varrhocsl\:\Esl(\Fst_{i} l_{j})
\end{equation}
in which the expected (mean) value
$\Esl(\cdot)$ is restricted to sliding contacts;
tangential forces $\Ft$ replace the full contact forces
$\mathbf{f}$; and
$\varrhocsl$ is the number of sliding contacts per volume.
\par
In this first attempt, we construct a strain-like $\Alphastruct$
by starting with the Liao--Chang estimate of the
full strain rate,
\begin{equation}
  \dot{\varepsilon}_{ij}\approx \dot{\varepsilon}_{ij}^{\text{LC}}
  =
  \left(
  E^{\text{p}}(l_{j} l_{k})
  \right)^{-1}
  E^{\text{p}}(\dot{l}_i l_{k})
\end{equation}
where $\dot{\mathbf{l}}$ is the relative velocity of
two particles' centers.
Here, $E^{\text{p}}(\cdot)$ is a mean applied to
\emph{pairs of contacting particles},
rather than to contacts,
noting that a pair of non-convex particles can have multiple
contacts.
The rate $\Varepsilondot^{\text{LC}}$ is known to be
inexact \cite{Bagi:2006a,Duran:2010b}
and, in the author's experience, underestimates
the actual strain rate
(measured by boundary movements)
by 5--20\%.
The attempted $\boldsymbol{\alpha}^{\text{str,?}}$
adapts the Liao--Chang strain to the slip velocities
$\Deltaslip$ of \emph{sliding contacts}
(again, the ``?'' denotes a tentative status),
\begin{equation}\label{eq:alphaq}
  \dot{\alpha}^{\text{str,?}}_{ij}
  =
  \left(
  \Esl(l_{j} l_{k})
  \right)^{-1}
  \Esl\big(\deltaslip_{i}l_{k}\big)
\end{equation}
using the expectation $\Esl(\cdot)$ for the subset of
sliding contacts.
\par
Unfortunately, the product
$\boldsymbol{\chi}^{\text{str,?}}\cdot\dot{\boldsymbol{\alpha}}^{\text{str,?}}$
is an unsuitable measure of dissipation,
as it is occasionally negative.
The reason is revealed with a derivation of the product.
By applying the product rule of expectations twice
to the true contact dissipations $\Fst_{i}\deltaslip_{i}$, we obtain
\begin{equation}\label{eq:attempt}
\begin{aligned}
  \chi^{\text{str,?}}_{ij}
  \dot{\alpha}^{\text{str,?}}_{ij}
  =\;&
  \varrhocsl\:\Esl(\Fst_{i} \deltaslip_{i})
  \\
  &+
  \varrhocsl\:\Esl(l_{j} l_{k})
  \:
  \left[
    \text{cov}
    (\Fst_{i}\deltaslip_{i},l_{j}l_{k})
    -
    \text{cov}
    (\Fst_{i} l_{j}, \deltaslip_{i} l_{k})
  \right]
\end{aligned}
\end{equation}
in which $\text{cov}(\cdot,\cdot)$ is the covariance of its two
quantities.
The first term on the right
is the true dissipation rate
(see Eq.~\ref{eq:DEMdissip}).
This dissipation is offset by two covariances,
with each arising from correlations between their two quantities.
Specifically,
dissipation $\Fst_{i}\deltaslip_{i}$
is known to preferentially occur among contacts of
certain orientations $l_{j}l_{k}$
\cite{Kuhn:2004k} (the first covariance),
and greater slip $\deltaslip_{i} l_{k}$ is known to occur
among more lightly loaded contacts $\Fst_{i} l_{j}$
\cite{Radjai:1998a,Kruyt:2007a} (the second covariance).
A yet more significant cause for the large, dominating second covariance
is that $\Ft$ is always aligned with slip
$\Deltaslip$~--- an unquestionable correlation~---
so that vectors can individually cancel at a given
orientation $\mathbf{l}$ even though they all contribute to
dissipation.
\par
The author considered other candidates for the
pair of $\Alphastruct$ and $\Chistruct$ with similar
results: occasional negative products
$\Alphastruct\cdot\Chistruct$.
The adopted candidate that we adopted
is one based on the alignment of the two contact vectors,
$\Ft$ and $\Deltaslip$, such that
$\Fst_{i}\deltaslip_{i}=|\Ft||\Deltaslip|$.
(This alignment applies to the linear-frictional contact model
used in our DEM simulations, although it might not
apply with other contact models.
E.g., \cite{Jager:2005a}).
The product rule of expectations yields the dissipation
\begin{equation}\label{eq:pre1}
\begin{aligned}
  \Alphastruct\cdot\Chistruct
  &=
  \varrhocsl\:
  \Esl\big(|\Ft||\Deltaslip|\big)
  \\
  &=
  \varrhocsl\:
  \Esl\big(|\Ft|\big)
  \Esl\big(|\Deltaslip|\big)
  +
  \varrhocsl\:
  \text{cov}
  \big(
    |\Ft|, |\Deltaslip|
  \big)
\end{aligned}
\end{equation}
We found that
the magnitudes $|\Ft|$ and $|\Deltaslip|$, although they are correlated,
have a covariance that is typically smaller than the
combined covariances of the previous candidate in Eq.~(\ref{eq:attempt}).
\par
The dissipation is estimated from
the product of the two scalars $\Esl(|\Ft|)$
and $\Esl(|\Deltaslip|)$, seen in the first term on the second
line of Eq.~(\ref{eq:pre1}).
Directional,
tensor forms of $\Alphastruct$ and $\Chistruct$
are motivated by Eqs.~(\ref{eq:chiq}) and~(\ref{eq:alphaq}),
which introduce the unit direction $\mathbf{t}^{\text{sl}}$ of a
contact's sliding
friction, with
$\mathbf{t}^{\text{sl}}=\Ft/|\Ft|=\Deltaslip/|\Deltaslip|$,
so that $\Fst_{i}=t^{\text{sl}}_{i}|\Ft|$
and $\deltaslip_{i}=t^{\text{sl}}_{i}|\Deltaslip|$.
Equation~(\ref{eq:pre1}) is multiplied by (and divided by)
the tensor $Q_{ij}$, thus maintaining equality:
\begin{align}\label{eq:pres}
&\begin{aligned}
  \Alphastruct\!\cdot\!\Chistruct
  =
  &\left(
    \frac{\varrhocsl}{\text{tr}(\boldsymbol{Q})}
    \Esl \big( |\Ft| \big)
    \Esl \big( t^{\text{sl}}_{i}l_{k}\big)
  \right)
  \left(
    \Esl\big( |\Deltaslip| \big)
    \Big( \Esl\big(l_{l} l_{k}\big)\Big)^{-1}
    \Esl\big( t^{\text{sl}}_{i} l_{l}\big)
  \right)
  \\
  &+
  \varrhocsl\:
  \text{cov}\big(|\Ft|, |\Deltaslip|\big)
\end{aligned}
\\
  \label{eq:Q}
  &Q_{ij}
  =
  \Esl\big( t^{\text{sl}}_{i} l_{k}\big)
  \left(
    \Esl\big( l_{l} l_{k} \big)
  \right)^{-1}
  \Esl\big( t^{\text{sl}}_{j} l_{l}\big)
\end{align}
The right side of the first row of Eq.~(\ref{eq:pres})
is the product of two precursor tensors,
designated $\boldsymbol{\chi}^{\text{pre}}$
and $\dot{\boldsymbol{\alpha}}^{\text{pre}}$.
This product is consistently positive
(in part, because the covariance in \ref{eq:pres}
is usually negative).
In Section~\ref{sec:micro2},
tensor $\Chistruct$ is taken as equal to
$\boldsymbol{\chi}^{\text{pre}}$,
whereas $\dot{\boldsymbol{\alpha}}^{\text{pre}}$
is scaled so that, without the covariance in Eq.~\ref{eq:pres},
the product $\Alphastruct\cdot\Chistruct$
equals the true dissipation 
$\varrhocsl\:\Esl\big(|\Ft||\Deltaslip|\big)$. 
\small
\bibliographystyle{elsarticle-num}

\end{document}